\documentclass[11pt]{article}
\usepackage[letterpaper,includefoot,top=1in,bottom=1in,left=1in,right=1in]{geometry}
\usepackage[centertags,sumlimits]{amsmath}
\usepackage{amsfonts}
\usepackage{amssymb,graphics,psfrag}
\usepackage{array,epsfig,multirow,stmaryrd,graphicx}
\usepackage[all,cmtip]{xy}
\usepackage{mathrsfs}
\usepackage[bookmarks,breaklinks]{hyperref}
\usepackage{multirow}
\usepackage[bookmarks,breaklinks]{hyperref}
\usepackage{amsthm}
\usepackage{color}

\numberwithin{equation}{section}
\numberwithin{table}{section}

\renewcommand{\Re}{\operatorname{Re}}

\newcommand{\beq}{\begin{equation}}
\newcommand{\eeq}{\end{equation}}
\newcommand{\be}{\begin{equation}}
\newcommand{\ee}{\end{equation}}
\newcommand{\bea}{\begin{eqnarray}}
\newcommand{\eea}{\end{eqnarray}}   
\newcommand{\ben}{\begin{eqnarray*}}
\newcommand{\een}{\end{eqnarray*}}                  
\newcommand{\ba}{\begin{aligned}}
\newcommand{\ea}{\end{aligned}}
\newcommand{\bt}{\begin{tabular}}
\newcommand{\et}{\end{tabular}}
\newcommand{\bc}{\begin{center}}
\newcommand{\ec}{\end{center}}
\newcommand{\cO}{\mathcal{O}}

\newcommand{\cC}{\mathcal{C}}
\newcommand{\cD}{\mathcal{D}}

\newcommand{\cK}{\mathcal{K}}
\newcommand{\cN}{\mathcal{N}}

\newcommand{\cG}{\mathcal{G}}
\newcommand{\cA}{\mathcal{A}}
\newcommand{\cH}{\mathcal{H}}
\newcommand{\cB}{\mathcal{B}}
\newcommand{\cF}{\mathcal{F}}

\newcommand{\cV}{\mathcal{V}}

\newcommand{\cM}{\mathcal M}

\newcommand{\I}{\text{Im}}
\newcommand{\R}{\text{Re}}

\newcommand{\bbZ}{\mathbb{Z}}
\newcommand{\bbR}{\mathbb{R}}

\newcommand{\nn}{\nonumber}

\newcommand{\cref}{{\bf [check ref]}}

\def\P{\mathbb{P}}

\entrymodifiers={+!!<0pt,\fontdimen22\textfont2>}

\begin{document}

\baselineskip=16pt
\setlength{\parskip}{6pt}

\begin{titlepage}
\begin{flushright}
\parbox[t]{1.4in}{
\flushright MPP-2012-5\\
UPR-1237-T\\
BONN-TH-2012-02}
\end{flushright}

\begin{center}

\vspace*{0.5cm}

{\Large \bf Fluxes and Warping for Gauge Couplings in F-theory
}

\vskip 1cm

\begin{center}
        \normalsize \bf{Thomas W.~Grimm}$^a$ \footnote{\texttt{grimm@mppmu.mpg.de}}, \normalsize 
        \bf{Denis Klevers}$^b$ \footnote{\texttt{klevers@sas.upenn.edu}}, \normalsize \bf{Maximilian Poretschkin}$^c$ 
        \footnote{\texttt{poretschkin@th.physik.uni-bonn.de}}
\end{center}
\vskip 0.5cm

 \emph{$^{a}$ Max Planck Institute for Physics, \\ 
                       F\"ohringer Ring 6, 80805 Munich, Germany} 
\\[0.25cm]
\emph{$^b$ Department of Physics and Astronomy,\\
University of Pennsylvania, Philadelphia, PA 19104-6396, USA}
\\[0.25cm]
\emph{$^c$ Bethe Center for Theoretical Physics, Universit\"at Bonn, \\
Nussallee 12, 53115 Bonn, Germany}
\\[0.15cm]
 \vspace*{0.5cm}

\end{center}

\vskip 0.0cm

\begin{center} {\bf ABSTRACT } \end{center}

We compute flux-dependent corrections 
in the four-dimensional F-theory effective action using 
the M-theory dual description. In M-theory the 
7-brane fluxes are encoded by four-form flux and
modify the background geometry and Kaluza-Klein reduction ansatz. In particular, the flux 
sources a warp factor which also depends on the torus directions 
of the compactification fourfold.  This dependence is crucial in the derivation of the four-dimensional 
action, although the torus fiber is auxiliary in F-theory. 
In M-theory the 7-branes are described by an infinite array of Taub-NUT spaces.
We use the explicit metric on this geometry to derive the locally corrected
warp factor and M-theory three-from as closed expressions.
We focus on contributions to the 7-brane gauge coupling function
from this M-theory back-reaction and show that terms quadratic in the 
internal seven-brane flux are induced.
The real part of the gauge coupling function is modified by the M-theory 
warp factor while the imaginary part is corrected due to 
a modified M-theory three-form potential.
The obtained contributions match the known weak string coupling result, but 
also yield additional terms suppressed at weak coupling.
This shows that the completion of the M-theory reduction
opens the way to compute various corrections
in a genuine F-theory setting away from the weak string coupling limit.

%\hfill \today
\end{titlepage}

%%%%%%%%%%%%%%%%%%%%%%%%%%%%%%%%%%%%%%%%%%%%%%%%%%%%%%%%%%%%%%%%%%%%%%%%%%%%%%%
%%%%%%%%%%%%%%%%%%%%%%%%%%%%%%%%%%%%%%%%%%%%%%%%%%%%%%%%%%%%%%%%%%%%%%%%%%%%%%%
\section{Introduction}
%%%%%%%%%%%%%%%%%%%%%%%%%%%%%%%%%%%%%%%%%%%%%%%%%%%%%%%%%%%%%%%%%%%%%%%%%%%%%%%
%%%%%%%%%%%%%%%%%%%%%%%%%%%%%%%%%%%%%%%%%%%%%%%%%%%%%%%%%%%%%%%%%%%%%%%%%%%%%%%

In Type IIB string theories non-Abelian gauge theories  with gauge groups including ADE 
can arise on stacks of 7-branes. 
These eight-dimensional  
branes are special, since they back-react on the geometry 
also far away from the brane with a non-trivial deficit angle.
F-theory provides a powerful non-perturbative description of 7-brane configurations
including their back-reaction on the geometry and the non-trivial dilaton-axion 
profile \cite{Vafa:1996xn,Denef:2008wq}. 
In an F-theory background the 
varying dilaton-axion is interpreted as the complex structure modulus
of an  auxiliary two-torus varying over the  ten-dimensional space-time.
This is the elliptic fibration of F-theory with 7-branes located 
at degeneration points of the torus fiber.
In four-dimensional compactifications on a complex threefold $B_3$ the inclusion
of a holomorphically varying dilaton-axion requires to consider an elliptically 
fibered Calabi-Yau fourfold $Y_4$ with a section. In this 
geometry the 7-branes wrap four-dimensional cycles  
in $B_3$ arising as the 
discriminant locus of the elliptic fibration of $Y_4$. 
The Calabi-Yau condition on $Y_4$ relates the class of the discriminant to the curvature
of $B_3$. This ensures that F-theory on $Y_4$  is canceling 7-brane tadpoles and 
takes the 7-brane back-reaction into account.

To study the lower-dimensional effective actions of 
F-theory compactifications it  is necessary to use 
the formulation of F-theory as an M-theory compactification in a special limit 
\cite{Denef:2008wq,Dasgupta:1999ss,arXiv:1008.4133}. 
This different point of view is inevitable due to the fact that F-theory 
has neither a low-energy effective action in twelve dimensions nor a fundamental 
microscopic formulation due to the strong coupling regions in a generic background. 
However, one argues that M-theory compactified on $Y_4$ to three 
dimensions is considered as four-dimensional F-theory on the same $Y_4$ compactified on an additional circle. 
This can be understood by applying the equivalence of M-theory 
on a two-torus via T-duality to Type IIB on a circle. 
In order to study any coupling of the  four-dimensional effective action of 
F-theory it will be crucial to carefully follow it through this duality. Having a given
question for F-theory in mind our strategy will thus be to first formulate this question in the dual 
M-theory and then perform the duality to F-theory. 

The inclusion of background fluxes in F-theory compactifications is of crucial interest, for example, in moduli 
stabilization \cite{Denef:2008wq,Douglas:2006es}, 
or more recently in the generation of a chiral spectrum 
\cite{Donagi:2008ca,Beasley:2008dc,Braun:2011zm,Marsano:2011hv,Krause:2011xj,Grimm:2011fx}. 
Moreover, they are often essential to cancel D3-brane tadpoles  \cite{Denef:2008wq,Douglas:2006es}.
However, the understanding of fluxes would be incomplete when ignoring their back-reaction. In the dual M-theory 
picture the F-theory fluxes correspond to four-form flux $\cG_4$ and generalizations thereof.  
It is known that such fluxes source a non-trivial warp factor on the M-theory background \cite{Becker:1996gj}. 
Thus also in F-theory the effect of warping
has to be included in general. In addition to yielding the well-studied effect of the Type IIB warp factor from 
an M-theory perspective 
\cite{Dasgupta:1999ss,Giddings:2001yu,Giddings:2005ff,Burgess:2006mn,Shiu:2008ry,Douglas:2008jx,Martucci:2009sf}, 
it is in general unclear what the dependence of the warp factor 
$e^{3A/2}$ on the 
\textit{non-physical} directions of the auxiliary two-torus of $Y_4$ should map to in F-theory. Recall that in F-theory 
the torus directions are not part of the physical spacetime but keep track of the dilaton-axion. Similarly, also a
back-reaction of the flux $\mathcal{G}_4$ altering the Kaluza-Klein reduction ansatz for the M-theory
three-form $C_3$ by including a non-closed three-form $\beta$ that is the Chern-Simons 
form of the flux $\mathcal{G}_4=d\beta$ has to be interpreted carefully in the F-theory dual. 
In this work, we follow both the dependence of the warp factor $e^{3A/2}$ on the unphysical 
torus direction and the non-closed three-form $\beta$ through the duality to F-theory and find that they map to 
corrections to 7-brane gauge coupling function for special choices of flux $\mathcal{G}_4=\omega_i\wedge \cF^i$. 
In the F-theory language $\omega_i$ are the $(1,1)$-forms from blow-ups of singularities of the elliptic fibration 
of $Y_4$ over a divisor $S_{\rm b}$ in $B_3$  
wrapped by a stack of 7-branes and $\cF^i$ denote 7-brane fluxes on $S_{\rm b}$. The 
corresponding corrections of the 7-brane gauge coupling function are shown to precisely reproduce the known flux 
corrections to the D7-brane gauge-coupling in the weak coupling limit of F-theory  \cite{Jockers:2004yj}. These 
arise at weak coupling from the eight-dimensional couplings of the form $\text{Tr}(F^4)$ on the 7-branes when brane 
fluxes are included. 
However, the M-theory to F-theory lift provides a formalism to 
compute corrections valid even away from the weak 
coupling limit of F-theory. In our example we match for small $g_s$ the weak coupling contributions 
but also find further 
corrections suppressed by at least one power of $g_s$.

A deeper understanding of the corrections to the 7-brane 
gauge coupling function is 
of crucial importance both from a conceptional 
as well as phenomenological point of view. 
Since in four-dimensional $\cN=1$ theories 
the gauge coupling function is holomorphic in the chiral multiplets one expects 
that it is one of the $\cN=1$ data which should be computable in a controlled way.
In the effective four-dimensional theory the leading gauge coupling 
of the 7-brane gauge theory is simply given 
by the volume of the cycle wrapped by the brane. 
Direct computations of the brane flux corrections to the gauge couplings are possible in certain 7-brane 
configurations in F-theory on K3 with constant string coupling as demonstrated in \cite{Lerche:1998nx,Lerche:1998gz}. 
It was also shown these corrections can be related to purely geometric data of the 
K3 compactification space.  
There is an immediate phenomenological relevance of these corrections when building, for example, Grand
Unified models in F-theory 
\cite{Donagi:2008ca,Beasley:2008dc,Marsano:2009ym,Blumenhagen:2009yv,Chen:2010ts,Cvetic:2010rq,Camara:2011nj}. 
In fact, the flux-induced corrections to the 
gauge coupling functions can crucially alter their running spoiling unification 
\cite{Donagi:2008kj,Blumenhagen:2008aw}.

While the geometric framework presented in this work is applicable more 
generally, our main focus will be on the computation of the corrections to 
the gauge coupling function in a local geometry.
%e will argue that the geometric framework presented in this work to compute the 
%corrections to the gauge coupling functions should apply for a general F-theory compactification. 
In this geometry we can use the local Calabi-Yau metric and 
show that the back-reaction of the flux $\cG_4$ can be evaluated explicitly. 
%However, the treatment of the back-reaction of the flux $\cG_4$ on the geometry can not be conducted for
%a compact Calabi-Yau fourfold $Y_4$ since the knowledge of its Calabi-Yau metric were required. 
More specifically, we consider a local model $\mathcal{Y}_4$ that appropriately describes $Y_4$ in the vicinity 
of a stack of $k$ 7-branes on a complex surface $S_{\rm b}$ in $B_3$. We construct this geometry explicitly by 
following 7-branes through the M-theory/F-theory duality, where the 7-branes become part of the M-theory geometry.
As the first step to identify the corresponding M-theory geometry, 
we consider M-theory compactified on a Taub-NUT space with $k$ centers $TN_k$.
A Taub-NUT space is a circle 
(Hopf) fibration over flat $\mathbb{R}^3$ apart from the $k$ centers where the circle 
shrinks to zero size. Its metric takes the form of the Gibbons-Hawking ansatz that is 
specified by $k$ functions $V_I$ that are formally the potentials of point charges in three 
dimensions \cite{Gibbons:1979zt}. 
For small asymptotic circle radii this setup yields Type IIA string 
theory with 6-branes located at the $k$ centers in $\mathbb{R}^3$ and filling the remaining 
seven dimensions. The asymptotic Taub-NUT radius is identified with the radius of the 
A-circle $r_{A}$, such that the Type IIA limit is $r_{A} \rightarrow 0$.
One of the $\mathbb{R}^3$ directions we like to place on a further circle, called the 
B-circle, such that a T-duality along this circle yields a stack of $k$ Type IIB 7-branes. 
Instead of putting the 6-branes on a circle, we can equally treat this as an image charge
problem of infinitely many 6-branes along a line with a fixed spacing that defines the circle 
circumference $r_B$. The advantage of this prescription is that one immediately 
infers the corresponding M-theory setup by extending the multi-Taub-NUT geometry $TN_k$ 
to infinitely many centers $TN_k^\infty$.  
Following this logic we argue that the M-theory metric still takes the Gibbons-Hawking form 
where $V_I$ are the potentials of an infinite array of periodically repeating 
point charges along a line in $\mathbb{R}^3$. 
The metric is obtained by a Poisson resummation and involves defining functions 
already found in \cite{Ooguri:1996me,Gaiotto:2008cd} in another context. The 
resulting geometry describes a torus fibration over the normal space 
$\mathbb{R}^2$ to the line of monopoles with torus fiber pinching at the $k$ centers in $\mathbb{R}^2$. 
In particular we read off the profile of the dilaton-axion from the metric data of $TN_k^\infty$.
Upon compactifying on the complex surface $S_{\rm b}$ we get a three-dimensional theory 
from M-theory on $\mathcal{Y}_4=TN_k^\infty\times S_{\rm b}$.
We apply T-duality to dualize the 6-branes into 7-branes in F-theory compactified on a circle
to three dimensions.

Having defined the M-theory or dual F-theory geometry $\mathcal{Y}_4$, we can specify the
M-theory flux $\mathcal{G}_4$ explicitly. We switch on G-flux supported on the explicitly 
constructible normalizable two-forms of the Taub-NUT space $TN^\infty_k$. This flux will 
descend to a 6-brane flux $\cF^I$ with non-trivial instanton number on $S_{\rm b}$ that is mapped to 
a worldvolume flux on the dual 7-brane. 
We calculate explicitly the corrections of this flux to the effective three-dimensional 
6-brane gauge coupling. In the M-theory picture this is induced by the back-reaction of 
the flux $\cG_4$ on the warp factor $e^{3A/2}$ near the centers of the Taub-NUT
spaces and on the Kaluza-Klein ansatz for the three-form $C_3$. 
In particular we are able to solve the equation for the warp factor using 
some core features of the metric on $TN_k^\infty$.
The corrections are sharply localized at the positions of the D6-branes in limit $g_s\rightarrow 0$. 
At weak Type IIB string coupling we show explicitly that these 
new terms encode corrections to the four-dimensional D7-brane 
gauge coupling function which are linear in 
the dilaton-axion $\tau$ and quadratic in the worldvolume flux $\cF^2$.

The paper is organized as follows. We start with a review of the gauge coupling on a stack of D7-branes 
in Type IIB in section \ref{sec:TypeIIBRev}. As a crucial ingredient we present the  $\cN=1$ 
effective action in terms of linear multiplets in section \ref{CYorientifolds} that we compactify on a circle to 
three dimensions in section \ref{sec:dimreduction3d}. This is necessary to compare to three-dimensional 
M-theory obtained from the warped compactification on a fourfold with G-flux as reviewed in section 
\ref{sec:FtheoryBasics}. We embed $k$ D6-branes as multi-center Taub-NUT into 
M-theory in section \ref{BasicsTNk} and compactify on $S^1$ by constructing an infinite periodic array of 
multi-center Taub-NUT $TN_k^\infty$ in section \ref{basics_TNinfty}. In section \ref{sec:LeadingGaugeKin}
we turn to the calculation of the 7-brane gauge coupling. We start with a light review of F-theory as M-theory
in section \ref{limit}. Then we determine the leading gauge coupling on a general compact fourfold in section 
\ref{sec:LeadingPart}. We extend the M-theory reduction to include a back-reaction of the G-flux on the three-form 
reduction ansatz in section \ref{sec:warpedreduction}. With these preparations we derive the full flux-
corrected gauge coupling in section \ref{sec:CalcCorrections}. We first  
obtain the real part from the back-reaction of the G-flux on the warp factor that we determine as a 
closed expression on $TN_k^\infty$ and then for the imaginary part by taking into account the altered 
reduction ansatz for the three-form that we also explicitly determine. We present our conclusions in 
section \ref{sec:conclusion} and provide additional details, in particular on the linear multiplet formalism 
and the construction of $TN_k^\infty$, in four appendices \ref{app:conventions4dN1} to 
\ref{app:TNchaingeo}.

%%%%%%%%%%%%%%%%%%%%%%%%%%%%%%%%%%%%%%%%%%%%%%%%%%%%%%%%%%%%%%%%%%%%%%%%%%%%%%%
%%%%%%%%%%%%%%%%%%%%%%%%%%%%%%%%%%%%%%%%%%%%%%%%%%%%%%%%%%%%%%%%%%%%%%%%%%%%%%%
\section{Motivation: D7-brane gauge coupling function}
\label{sec:TypeIIBRev}
%%%%%%%%%%%%%%%%%%%%%%%%%%%%%%%%%%%%%%%%%%%%%%%%%%%%%%%%%%%%%%%%%%%%%%%%%%%%%%%
%%%%%%%%%%%%%%%%%%%%%%%%%%%%%%%%%%%%%%%%%%%%%%%%%%%%%%%%%%%%%%%%%%%%%%%%%%%%%%%

In this section we discuss the aspects of the four-dimensional effective theory 
on a stack of D7-branes in a weakly coupled orientifold 
compactification. We first recall in section \ref{CYorientifolds} the expression of the 
D7-brane gauge coupling function as determined by a reduction 
of the D7-brane action. In section \ref{sec:dimreduction3d} we perform the reduction to three dimensions. 
This three-dimensional 
result will be later useful to compare to the F-theory gauge 
coupling function derived via M-theory.

%%%%%%%%%%%%%%%%%%%%%%%%%%%%%%%%%%%%%%%%%%%%%%%%%%%%%%%%%%%%%%%%%%%%%%%%%%%%%%%
\subsection{D7-brane gauge couplings in 4d: Calabi-Yau orientifolds} \label{CYorientifolds}
%%%%%%%%%%%%%%%%%%%%%%%%%%%%%%%%%%%%%%%%%%%%%%%%%%%%%%%%%%%%%%%%%%%%%%%%%%%%%%%

We begin by recalling the basics from the computation of the four-dimensional 
D7-brane effective action in Type IIB $\cN=1$ compactifications on a 
Calabi-Yau  orientifold $B_3=Z_3/\sigma$ with 
O7-planes \cite{Grimm:2004uq,Jockers:2004yj}.\footnote{See \cite{Grimm:2008dq,Grimm:2011dx,Kerstan:2011dy} for 
a similar derivation of the dual D5- respectively D6-brane effective action.} 
Here $Z_3$ denotes the Calabi-Yau threefold covering space of the 
orientifold $B_3$ that is obtained by modding out an holomorphic 
involution $\sigma:\, Z_3\rightarrow Z_3$. For appropriately chosen involution the 
fix point locus of $\sigma$ is a holomorphic divisor $D_{O7}$ that supports 
the O7-orientifold planes. The divisor $D_{O7}$ has to be homologous to 
$-8$ times the divisors $S$ in $Z_3$ wrapped by the D7-branes due to 
tadpole cancellation.

String theory on the orientifold is specified by the orientifold action
$\cO$ = $\Omega (-1)^{F_L} \sigma$ acting on the fields, with $\Omega$ being the worldsheet parity operator, and 
$F_L$ the left-moving fermion number. The spectrum of orientifold invariant states, i.e.~states transforming with an 
eigenvalue 1, determines the physical spectrum.  
To obtain the four-dimensional effective theory, all massless fields both of the bulk and the D7-brane have to be 
expanded in zero-modes that are counted by appropriate cohomology groups. 
Then these expansions are inserted into the ten-dimensional Type IIB supergravity and eight-dimensional D7-brane 
effective action, that are dimensionally reduced to four dimensions by integration over the internal directions and 
keeping only the orientifold invariant terms, to obtain the $\mathcal{N}=1$ effective four-dimensional action.

Let us outline the gauge sector of this effective theory focusing on a stack of $k$ 
D7-branes with an eight-dimensional U$(k)$ gauge theory on $\mathbb{R}^{(3,1)}\times S_{\rm b}$. 
If the divisor $S_{\rm b}$ has a non-trivial topology, one can consider flux 
configurations $\cF$ for the field strength $F_{\rm D7}$ on the D7-brane.
More precisely, we split in the Kaluza-Klein ansatz the 
D7-brane field strength as
\beq \label{eq:FD7exp}
   F_{\rm  D7} = F + \cF = (F^0 + \cF^0 ) \mathbf{1}+(F^i + \cF^i ) T_i + (F^A + \cF^A ) \tilde T_A\ ,
\eeq
where $F = dA + A \wedge A$ is the U$(k)$ field strength in the four-dimensional 
effective theory, and $\cF$ is a background two-form flux on the 
D7-brane divisor $S_{\rm b}$. The field strength $F_{\rm  D7}$
is a general element in the adjoint of U$(k)$, 
that we have expanded in the generators $\tilde T_{\cA}=(\mathbf{1},T_i,\tilde T_A)$ of the adjoint 
Here $T_I=(T_i,\mathbf{1})$, $i=1,\ldots, k-1$, are the $k$ Cartan generators
of U$(k)=$U$(1)\times $SU$(k)$, while $\tilde T_A$ denote the generators associated to the roots of SU$(k)$.

In the absence of flux $\cF$ it can be shown by a straightforward reduction of 
the D7-brane worldvolume action using the expansion \eqref{eq:FD7exp} 
that the kinetic term for $F$ takes the form
\beq \label{4daction}
   S^{(4)}_{F_{\rm D7}^2} = -2\pi \int_{\cM_4} \tfrac12 \R f_{\cA \cB} \, F^\cA  \wedge * F^\cB+ \tfrac12 \I f_{\cA \cB} \, 
   F^\cA \wedge F^\cB\, 
\eeq
in the conventions \eqref{eq:unitsconventions} of appendix \ref{app:conventions4dN1}.
Here $f_{\cA \cB}$ denotes the gauge coupling function that is a holomorphic 
function of the chiral fields in the $\mathcal{N}=1$ effective theory, and has adjoint 
indices $\cA,\cB$. The adjoint indices arise as we will show soon from the two traces 
\beq \label{tracesC}
   \cC_{\cA \cB} =   \text{Tr}(\tilde T_{\cA} \tilde T_{\cB} ) \ , \qquad \tilde \cC_{\cA \cB} =  \tfrac{1}{2} \, \text{sTr}(\tilde T_{\cA} \tilde T_{\cB}\tilde T_{\cC} \tilde T_{\cD} ) \int_{S_{\rm b}} \cF^{\cC} \wedge \cF^{\cD}\ ,
\eeq
where $\cF^{\cC}$ are the fluxes localized on the internal part $S_{\rm b}$ of the D7-brane and sTr$(.)$ 
denotes the symmetrized trace defined as the sum over all permutations $\sigma$,
\beq
	\text{sTr}(\tilde T_{\cA}\tilde T_{\cB}\tilde T_{\cC}\tilde T_{\cD})=\frac{1}{4!}\sum_{\sigma}\text{Tr}(\tilde T_{\sigma(\cA)}\tilde T_{\sigma(\cB)}\tilde T_{\sigma(\cC)}\tilde T_{\sigma(\cD)})
\eeq 
In the case at hand the chiral superfields are given by the
the axiodilaton $\tau$ = $C_0$ $+$ $ie^{-\phi}$, the combination $G^a=\int_{\Sigma_a}C_2-\tau \mathcal{B}_2$ 
and the K\"ahler moduli \cite{Grimm:2004uq}
\beq \label{eq:KaehlerO7}
T_{\alpha}= \int_{D_{\alpha}} \tfrac{1}{2}( J \wedge J-e^{-\phi} B_2\wedge B_2)+ i (C_4-C_2\wedge B_2 +\tfrac{1}2 C_0 B_2\wedge B_2)\ ,
\eeq
where $\Sigma_a$, $D_\alpha$ denote a homology basis of odd curves respectively even divisors in $Z_3$ w.r.t.~the involution $\sigma$. 
The K\"ahler form on $Z_3$ is given by $J$, while $C_p$ denote the R-R $p$-forms, and $B_2$ is the NS-NS B-field. 
For simplicity we have frozen out the position and 
Wilson line moduli of the D7-brane and refer to \cite{Jockers:2004yj} for the open 
string corrected chiral coordinates.

In order to proceed we will need to recall some additional facts about the 
D7-brane theories following \cite{Jockers:2004yj,Grimm:2010ez,Grimm:2011tb}. In particular, 
one finds that in the weak coupling description the gauge group is actually U$(k)= $SU$(k) \times $U$(1)$.
However, if the D7-brane and its orientifold image are not in the same cohomology 
class on $Z_3$, one finds that a geometric St\"uckelberg term is induced which 
renders the overall U$(1)$ massive. More precisely, the moduli $G^a$ are 
gauged due to the geometric St\"uckelberg coupling, and $\R G^a$ is eaten by 
the overall U$(1)$ which thus becomes massive. The mass of the massive 
vector multiplet containing the U$(1)$ and $\I G^a$ is 
of the order of the Kaluza-Klein scale. In the following we will make the 
simplifying assumption, that for each stack of D7-branes there is exactly one 
$G^a$ which becomes massive together with the overall U$(1)$. While 
a detailed derivation of the effective action 
would require to actually integrate out this 
massive vector mulitplets, we will in the following mostly drop it 
in our consideration. In other words, we will consider an SU$(k)$
gauge theory and no $G^a$ moduli.

Given these preliminaries we are now in the position to display 
the gauge coupling function $f_{\cA \cB}$ for a stack of D7-branes. 
This generalizes the results given for a single D7-brane \cite{Jockers:2004yj}.
Using the traces \eqref{tracesC} one finds
 \footnote{We have set $2 \pi \alpha'=1$ in the following.}
\bea \label{eq:fIJD74d}
    f_{\cA \cB}  &=&   \tfrac14 \, (\delta^\alpha_S T_\alpha \cC_{\cA \cB} -i\tau \, \tilde \cC_{\cA \cB}) \\
   &\equiv&  f^{\rm c}(T)\, \cC_{\cA \cB} + f^{\rm flux}_{\cA \cB}(\tau) \  . \nn
\eea
where $\delta^\alpha_S$ are the coefficients in the expansion of $S = \delta^\alpha_S D_\alpha$ in a homology 
basis of orientifold-even divisors. Note that the general $\cN=1$ effective action \eqref{4daction} with the
gauge coupling \eqref{eq:fIJD74d} is not a standard $\cN=1$ action 
due to the presence of the flux correction in \eqref{eq:fIJD74d}. These fluxes actually break the gauge group 
in the eight-dimensional world volume theory of the D7-branes. To make this more 
explicit we display the action splitting into a flux-independent and a flux-dependent part as
\bea \label{action_split}
    S^{(4)}_{F_{\rm D7}^2} &=& -2\pi \int_{\cM_4} \tfrac12 \R f^c \, \text{Tr}(F  \wedge * F)+ \tfrac12 \I f^{c} \, \text{Tr}(F \wedge F) \\
                           &&    \phantom{-2\pi |_{\cM_4}}\qquad \quad \ + \tfrac12 \R f^{\rm flux}_{\cA \cB} \, F^\cA  \wedge * F^\cB+ \tfrac12 \I f^{\rm flux}_{\cA \cB} \, F^\cA \wedge F^\cB\ , \nn 
\eea
Clearly, a standard $\cN=1$ action can be found if the fluxes are zero and the gauge group is completely unbroken. 
A second possibility is to consider the breaking of the group, for example by moving
the D7-branes apart on $Z_3$. Then one finds a standard $\cN=1$ action for 
a gauge group U$(1)^k$.  For completeness we will summarize the result in this phase. 
Later on we will T-dualize the D7-branes to D6-branes which can then be moved apart
in the T-dualized direction. 

Assuming that we can move the D7-branes apart on different internal cycles in the same class $[S_{\rm b}]$. The
gauge coupling function can be given for each individual brane labeled by $I=1,\ldots,k$. Fluxes are now only 
located on each separate D7-brane, which is reflected in the structure of adjoint indices. Indeed, in evaluating 
$\cC_{IJ}$ and $\tilde{\cC}_{IJ}$ from \eqref{tracesC} we use the basis $E_I=\text{diag}(0,\ldots,0,1,0,\ldots,0)$ with 
$1$ at the $I$-the position that is related to the $T_I=(T_i,\mathbf{1})$ by a basis transformation. We then readily evaluate \eqref{tracesC} as
\beq \label{tracessingle}
	\cC_{IJ}=\delta_{IJ}\,,\qquad \tilde{\cC}_{IJ}=\frac{1}{2}\delta_{IJ}\delta_{KL}\delta_{IK}\int_{S_{\rm b}}\cF^K\wedge\cF^L=\frac{1}{2}\delta_{IJ}n^I\,,
\eeq
where we exploited that the $E_I$ commute to evaluate the symmetrized trace sTr$(.)$.
Here $\cF^I$ denotes the internal flux on the $I$th D7-brane. The numbers $n^I$ characterize the topology of the 
gauge configuration on the $I$-th brane. They are related to the integral instanton number $k^I$ of 
the U$(1)$ on the $I$-th brane as $n^I=-8\pi^2 k^I$. Using these results the 
gauge coupling function on the $I$-th D7-brane is given by
\beq \label{f_Isingle}
    f_{I}  =   \tfrac14 \,(\delta^\alpha_S T_\alpha - i\tfrac12\tau \, n^I) \ .
\eeq

As we will see in section \ref{sec:LeadingGaugeKin} for the comparison of the D7-brane action with the M-theory 
fourfold compactification  it turns out to be convenient to dualize certain scalars into 
form fields. More precisely, we replace in four dimensions  
the chiral multiplet containing the complex scalars $T_\alpha$ with a linear multiplet containing the 
bosonic fields $(L^\alpha, \cC^\alpha_2)$. Here $L^\alpha$ are real scalars dual to the real part $\R T_\alpha$ and 
the imaginary part $\I T_\alpha$ is dual due to its shift symmetry to a two-form $\cC_2^\alpha$. 
It will then be crucial to follow the terms involving $f_{\cA \cB}$ through the dualization. As outlined in detail in 
appendix \ref{app:linMultis+reduction} this procedure dualizes the classical coupling $\I f^{\rm c}(T)\text{Tr}(F\wedge 
F)$ in \eqref{action_split} into a modification of the field strength strength $\cH^\alpha_3$ of $\cC_2^\alpha$ by 
the Chern-Simons form $\omega_{\rm CS}$ to $\text{Tr}(F\wedge F)$,
\beq \label{eq:modifiedFieldStrengthH}
\cH^\alpha_3  =   
d\cC_2^{\alpha} +  \tfrac{1}{8} \delta_S^{\alpha} \omega_{\rm CS}\ , \qquad \omega_{\rm CS} = A \wedge dA + 
\tfrac23 A\wedge A \wedge A\,. 
\eeq
The complete dual action as given in \eqref{eq:C2action} of appendix \ref{app:linMultis+reduction} then 
contains all terms in \eqref{action_split} except the term involving $\I f^{\rm c}(T)$ which is replaced, 
together with the kinetic term for the $\I T_\alpha$, by a kinetic term for $\cH^\alpha$. Of 
course all other fields that do not couple to $\I T_\alpha$ like $\tau$ and $\R 
T_\alpha$ or its dual $L^\alpha$ are unaffected. For the later comparison to M-theory it is important to keep 
in mind the  K\"ahler potential $\tilde{K}$ for $L^\alpha$ and $\tau$ obtained by Legendre transformation 
of $K(\tau\vert T)$ as
\begin{equation}
	\tilde{K}(\tau|L)=K+L^\alpha\, \R T_\alpha =\log(\tfrac{1}{6}L^\alpha L^\beta 
	L^\gamma\mathcal{K}_{\alpha\beta\gamma})-\log(\tau-\bar{\tau})\,.
\label{eq:LegendreKaehlerpotIIBMainText}
\end{equation}

%%%%%%%%%%%%%%%%%%%%%%%%%%%%%%%%%%%%%%%%%%%%%%%%%%%%%%%%%%%%%%%%%%%%%%%%%%%%%%%
\subsection{Dimensional reduction to three dimensions}
\label{sec:dimreduction3d}
%%%%%%%%%%%%%%%%%%%%%%%%%%%%%%%%%%%%%%%%%%%%%%%%%%%%%%%%%%%%%%%%%%%%%%%%%%%%%%%

In this subsection we discuss the circle reduction of the four-dimensional effective action of a D7-brane in an 
orientifold compactification to three dimensions. The final result will later be compared to the M-theory 
reduction on a Calabi-Yau fourfold when restricted to the weak coupling limit. It is important to stress that 
the M-theory reduction is performed on a smooth geometry at large volume. In the three-dimensional effective 
theory this yields a gauge theory on the Coulomb branch. In 4d F-theory compactified on an extra circle
new terms in the effective theory are generated due to the necessity to integrate out 
massive vector multiplets containing the W-bosons and charged chiral matter 
multiplets \cite{Grimm:2011fx,Bonetti:2011mw}.

In the D7-brane picture we 
consider a reduction on a circle of circumference $r$. 
Moving on the Coulomb branch is achieved by giving the scalars in the three-dimensional 
vector multiplets a vacuum expectation value.
In order to make this more precise 
we make the following reduction ansatz for the four-dimensional fields, 
\beq \label{KKa}
g_{\mu \nu}^4 = \begin{pmatrix} g_{pq} + r^2  A^0_{q} A^0_p & r^2 A^0_{q} \\ r^2 A^0_{p} & r^2 \end{pmatrix}\,, 
\qquad A= (A_3 - A^0\zeta,\zeta) \,. 
\eeq
Here $A_3$ and $\zeta$ are a three-dimensional vector and a three-dimensional scalar both transforming in the 
adjoint of the gauge group $G$. The Coulomb branch is obtained by giving $\zeta$ a vev, and splitting 
\beq
   \text{U}(k)\ \rightarrow \ \text{U}(1)^{k}\ , \qquad \quad A \ \rightarrow \ A^I \ , \qquad \quad \zeta \ 
   \rightarrow \ \zeta^I\ ,
\eeq
where $I = 1,...,k$ runs only over the Cartan generators $T_I$ of $\text{U}(k)$. 
In this split one can now evaluate the traces \eqref{tracesC}. By the basis change to the $E_I=\text{diag}(0,
\ldots,1,\ldots,0)$ the traces can be written, by the same calculations leading to \eqref{tracessingle}, as
\beq \label{eq:tracesCoulomb}
   \cC_{IJ} = \delta_{IJ}\ ,\qquad \tilde \cC_{IJ} = \tfrac{1}{2} \delta_{IJ} n^I\ ,
\eeq
where we used the numbers $n^I$ introduced before.
The couplings of the gauge-fields are thus encoded by 
\beq \label{eq:f_IJCoulomb}
  f_{IJ} =\tfrac14( C^{\alpha}_{IJ} T_{\alpha} - i\tfrac12 \tau \delta_{IJ} n^I)\ , \qquad C^\alpha_{IJ} = 
  \delta^\alpha_S \cC_{IJ}\ .
\eeq
Note that this breaking has a natural interpretation in the T-dual picture, where the T-duality is performed 
along the reduction circle. In this duality the D7-branes become D6-branes localized on points of the reduction 
circle. The Coulomb branch corresponds to moving the D6-branes apart. The $\zeta^I$ can then be reinterpreted 
as positions on the circle.

Since we are reducing an $\mathcal{N}=1$ supersymmetric action in four dimensions, we obtain an action with
$\cN$ = 2 supersymmetry in three dimensions. It can be brought into the form 
\bea \label{S3}
S^{(3)} = 2\pi\int_{\cM_3}& -\frac{1}{2}R_3 * 1 - \tilde K_{a \bar b}dM^a \wedge * d \bar M^{\bar b} + \frac{1}
{4} \tilde K_{\Lambda\Sigma} d\xi^{\Lambda} \wedge * d\xi^{\Sigma} \nn \\ & + \frac{1}{4} \tilde K_{\Lambda 
\Sigma} F^{\Lambda} \wedge * F^{\Sigma} + F^{\Lambda} \wedge \I (\tilde K_{\Lambda a} dM^a),
\eea
where one has to perform the Weyl-rescaling $g_{pq} \rightarrow  r^2 g_{pq}$ to the Einstein-frame metric, and 
to make the following identifications,
\beq 
\label{eq:3dIdentification}
R = r^{-2}\,, \qquad \xi^{\Lambda} = (R, R\zeta^{I})\,, \qquad A^{\Lambda} = (A^0, A^{I})\,,
\eeq
with $\Lambda=0,1,\ldots, k$.
Here, the $M^a$ collectively denote four-dimensional chiral multiplets.
The three-dimensional kinetic potential $\tilde K$ depends on $M^a$, $\bar{M}^{\bar b}$ as well as 
$\xi^{\Lambda}$ and reads
\beq \label{Kahler3}
\tilde K = K(M, \bar M) + \log R - \frac{1}{R}\R f_{IJ}(M)\xi^{I}\xi^{J}\ ,
\eeq
where $K(M,\bar M)$ is the four-dimensional K\"ahler potential evaluated for the three-dimensional fields. 
This kinetic potential contains also the gauge kinetic function since the third component of the 
four-dimensional vectors have become scalars $\xi^I$ in three dimensions as is obvious from \eqref{KKa}.

Let us be more concrete by reducing the four-dimensional action \eqref{eq:C2action} with linear multiplets.
Since we are considering not only chiral and vector multiplets, but also linear multiplets containing 
$\cC_2^{\alpha}$, the form of the kinetic action in three dimensions we will obtain 
will be different from \eqref{S3}.\footnote{However, we note that an action including linear multiplets can be 
brought in the standard form using duality of vectors and scalars in three dimensions.} Incorporating two-forms 
we specify the reduction ansatz for $\cC_2^{\alpha} $ such that
\beq
 \cH^\alpha_3 \quad \rightarrow\quad (\tilde{F}^{\alpha} + \tfrac14 C^{\alpha}_{IJ}\zeta^{I}F^J) \wedge dy\,,
\eeq
where we introduced the field strength $\tilde{F}^{\alpha}$ = $d \cA^{\alpha}$ in three dimensions. Physically this 
means that the fields $T_\alpha$, of which 
$\I T_S = \delta^\alpha_S T_\alpha$ constitutes the leading part of the D7-brane gauge coupling, 
will occur after dualization into two-forms and dimensional reduction as vectors in three dimensions. 
Plugging that into the action \eqref{eq:C2action} and performing a Weyl rescaling $g_{\mu \nu}$ $\rightarrow$ 
$r^2 g_{\mu \nu}$ we integrate out the circle coordinate $y$ to obtain  
\bea \label{eq:C2actionred}
S^{(3)}_{F^I,F^\alpha}&=&2\pi\int_{\cM_3} \tilde K_{\alpha \beta}  (\tilde{F}^{\alpha} +  \tfrac14 C^{\alpha}_{IJ} \zeta^{I} F^J) 
         \wedge * (\tilde{F}^{\beta} +  \tfrac14  C^{\beta}_{IJ} \zeta^{I} F^J)\nn \\ 
&&\qquad \qquad -\tfrac1{2R}\R f_{IJ} (F^{I} \wedge * F^J + d \xi^{I} \wedge * d \xi^J ) -  \I f^{\rm flux}_{IJ}  d\zeta^{I} \wedge F^J\,, % +  \tfrac{1}{4} \tilde K^{\alpha \beta} d \R T_{\alpha} \wedge * d \R T_{\beta} 
%- \tilde K_{\tau \overline{\tau}} d\tau \wedge * d\overline{\tau}\  . 
\eea
where we used the kinetic potential \eqref{Kahler3} with the four-dimensional K\"ahler potential 
\eqref{eq:LegendreKaehlerpotIIBMainText}.
As one can easily check this matches the structure anticipated in \eqref{S3} which is supplemented by 
additional terms involving the vectors $\tilde{F}^\alpha$ contributed by the linear multiplets. The terms to 
determine $f_{IJ}$ are: 
\vspace*{-.3cm}
\begin{itemize} 
\item[(1)] the 
kinetic term $F^I \wedge *F^J$ to determine the complete $\R f_{IJ}$, 
\item[(2)] the mixed terms $F^I \wedge * \tilde F^\alpha$
to determine the classical part of $\I f_{IJ}$ proportional to $\I T_\alpha$,
\item[(3)] the term $d\zeta^{I} \wedge F^J$ to obtain 
$\I f^{\rm flux}_{IJ} $.
\end{itemize}

%Thus, we conclude that from the point of view of the three-dimensional $\cN=2$ effective action that the leading term of the imaginary part of the gauge coupling function should be read off from the following term in the Type IIB or M-theory compactification:
%\beq
%\pi\int_{\cM_3}  \tilde K_{\alpha \beta} \tilde{F}^{\alpha} \wedge * C^{\beta}_{ij} \zeta^{i} F^j\,.
%\eeq
%In contrast the corrections $h_\alpha(\tau)$ can be read off from the terms
%\beq \label{eq:correctionterm}
%2\pi\int_{\cM_3} \tilde{F}^{\alpha} \wedge d\I h_{\alpha}, \quad \text{or} \quad \frac{2\pi}{4}\int_{\cM_3} C^{\alpha}_{ij} \zeta^{i} F^j \wedge d \I h_{\alpha}\,.
%\eeq
%Note that the first of these two terms is a total derivative, if $\tilde{F}$ is closed. 

Let us close this section by commenting on another choice of Cartan generators for $\text{U}(k)$ which 
naturally appears in M-theory.
This choice is associated to the split $\text{U}(k)=\text{SU}(k)\times \text{U}(1)$ and yields the trace $\cC_{IJ}$ in \eqref{tracesC} as
\beq
  \cC_{ij}=C_{ij}, \qquad \cC_{00} = k \ , \qquad \cC_{i0} = 0\ ,
\eeq
where $i,j = 1,...,k-1$ label the Cartan generators $T_i=E_i-E_{i+1}$ of SU$(k)$ and $C_{ij}$ is the Cartan matrix of SU$(k)$. Decoupling the overall U$(1)$ of $\tilde{T}_0=\mathbf{1}$ in $\text{U}(k)$ as in 
\cite{Grimm:2010ez,Grimm:2011tb}, the classical part of the three-dimensional 
gauge coupling function \eqref{eq:fIJD74d} splits for the Cartan U$(1)$'s of SU$(k)$ as 
\beq \label{eq:fIJD7}
    f_{ij} =  \tfrac{1}{4}\, C^{\alpha}_{ij} \, T_\alpha \  , \qquad \qquad C_{ij}^\alpha = C_{ij} \delta_S^\alpha\ .
\eeq
It was this coupling which was found in \cite{arXiv:1008.4133} in a dimensional reduction of M-theory on a 
resolved Calabi-Yau fourfold. We will recall this reduction briefly in section \ref{sec:LeadingPart}.

%%%%%%%%%%%%%%%%%%%%%%%%%%%%%%%%%%%%%%%%%%%%%%%%%%%%%%%%%%%%%%%%%%%%%%%%%%%%%%%
%%%%%%%%%%%%%%%%%%%%%%%%%%%%%%%%%%%%%%%%%%%%%%%%%%%%%%%%%%%%%%%%%%%%%%%%%%%%%%%
\section{M-theory compactifications and Taub-NUT geometries} \label{M-theory_geom}
%%%%%%%%%%%%%%%%%%%%%%%%%%%%%%%%%%%%%%%%%%%%%%%%%%%%%%%%%%%%%%%%%%%%%%%%%%%%%%%
%%%%%%%%%%%%%%%%%%%%%%%%%%%%%%%%%%%%%%%%%%%%%%%%%%%%%%%%%%%%%%%%%%%%%%%%%%%%%%%

In order to understand the  
gauge kinetic function of 7-branes in F-theory, 
we have to extend the Type IIB effective action discussed in the last section 
away from the weak coupling limit. This is achieved by considering F-theory 
as a limit of M-theory with G-fluxes.  
This section provides  the necessary background material from the M-theory 
perspective to determine the full gauge kinetic coupling function of 7-branes from 
back-reaction effects as demonstrated in section \ref{sec:LeadingGaugeKin}. 

In order to set the stage we first introduce the M-theory backgrounds 
with a non-trivial four-form flux $G_4$ in subsection \ref{sec:FtheoryBasics}.
In particular, we stress that the $G_4$ background induces a non-trivial warp 
factor. Later on this back-reaction will be shown to correct the gauge coupling 
function. Since we
will be interested in the gauge dynamics of one stack of 7-branes 
it will be necessary to introduce the dual local M-theory geometries.
For a stack of $k$ D7-branes the form of this local M-theory geometry 
can be inferred via string duality. First we note that in compactifying 
Type IIB on a circle one can T-dualize the D7-branes into $k$ D6-branes.
These D6-branes lift in  M-theory to the geometry of Kaluza-Klein 
monopoles. Since the metric and cohomology of Kaluza-Klein monopoles in M-theory
is just given by Taub-NUT\footnote{This name is due to Taub 
and Newman, Unti and Tamburino (NUT), but can also be traced back to nut at the origin
which is the terminus for an isometrical fixed point introduced by Hawking. } 
space $TN_k$ with $k$ indicating the number of monopoles, we can explicitly 
analyze their local geometry in subsection \ref{BasicsTNk}.

Having introduced the multi-Taub-NUT spaces we discuss in subsection \ref{basics_TNinfty}
a further compactification on a circle on which one can perform a 
T-duality to the F-theory setup. The resulting geometry will serve as 
a local model of the singular elliptic fibration of
the M-/F-theory fourfold $Y_4$ with a 7-brane located on a 
divisor $S_{\rm b}$ in the base. 
The compactification of the Taub-NUT geometry is achieved by 
considering an infinite chain of Kaluza-Klein monopoles with period $a$, denoted $TN_k^\infty$, 
and later considering the quotient. 
Technically, this process involves a resummation of certain divergent 
infinite sums in the corresponding metric.

%%%%%%%%%%%%%%%%%%%%%%%%%%%%%%%%%%%%%%%%%%%%%%%%%%%%%%%%%%%%%%%%%%%%%%%%%%%%%%%
\subsection{M-theory on warped Calabi-Yau fourfolds}
\label{sec:FtheoryBasics}
%%%%%%%%%%%%%%%%%%%%%%%%%%%%%%%%%%%%%%%%%%%%%%%%%%%%%%%%%%%%%%%%%%%%%%%%%%%%%%%

In this section we introduce vacuum solutions of M-theory on Calabi-Yau fourfolds 
with background fluxes following \cite{Becker:1996gj}.  
The eleven-dimensional low effective energy action of M-theory is given by \footnote{We have set $\ell_{\rm M} = 1$ in the conventions of \cite{Denef:2008wq}.}
\beq \label{Mtheoryaction}
S^{(11)}_{\rm M} = -2\pi \int_{\cM_{11}}  \tfrac12 R * 1 
+ \tfrac{1}{4} G_4\wedge * G_4 +\tfrac{1}{12}C_3\wedge G_4\wedge G_4  
- 2\pi \int_{\cM_{11}} C_3\wedge X_8 + \sum_k S^k_{\rm M2}
\eeq
where locally $G_4= dC_3$ is the field strength of the M-theory 
three-form $C_3$, and $X_8$ is a forth order polynomial in the Riemann 
curvature of the eleven-dimensional space-time. 
The last term includes the coupling to M2-branes with action $S^k_{\rm M2}$.
The $G_4$ field strength, the curvature $X_8$, and the presence of M2-brane
can serve as sources in the $C_3$ equations of motion 
\beq
\label{eq:BianchiId}
   d * G_4 =\tfrac12 G_4 \wedge G_4 - X_8 + \sum_k \delta^{(8)}({\Sigma^k_3})\, ,
\eeq
where $\delta^{(8)}({\Sigma^k_3})$ is an eight-form current localizing on the world-volumes
$\Sigma_3^k$ of the M2-branes.

Supersymmetric solutions can be analyzed by solving 
the equations of motion of \eqref{Mtheoryaction} and its supersymmetry 
variations. A non-trivial background was found in  \cite{Becker:1996gj}
which allows for an internal Calabi-Yau geometry $Y_4$ times a 
flat space $\mathbb{R}^{(2,1)}$ and a background flux 
for the field strength $G_4$.
The metric in the presence of such flux has to include a non-trivial 
warp factor $e^A$, and is given by
\begin{equation} \label{warped-background}
 ds^2_{(11)} = e^{-A}\eta_{\mu \nu} dx^\mu dx^\nu + e^{A/2} g_{a \bar b } dy^{a} d\bar y^{ \bar{b}}\ ,
\end{equation}
with $g_{a \bar b}$ being the metric on the Calabi-Yau manifold 
$Y_4$. The warp factor only depends on the coordinates $y^a,\bar y^{b}$ of $Y_4$.
The non-trivial field strength $G_4$ splits into a 
contribution with three flat indices $(G_4)_{\mu \nu \rho m} $
and an internal $G_4$-flux $\cG_4$ with indices only 
along $Y_4$.
Supersymmetry implies the background component of $G_4$ with flat indices is 
determined by the warp factor
\beq \label{G4flat}
(G_4)_{\mu \nu \rho m} = \epsilon_{\mu \nu \rho} \partial_m e^{3A/2}\,,
\eeq
where the derivative is taken with respect to the internal coordinates.
The equations \eqref{G4flat} and \eqref{eq:BianchiId} require that the 
warp factor has to fulfill the Laplace equation
\begin{equation} \label{Laplaceeq}
 \Delta_{Y_4} (e^{3A/2}) = *_{Y_4} (\tfrac12\cG_4 \wedge \cG_4 - X_8|_{Y_4}  +\sum_k \delta^{(8)}
 ({\Sigma^k_3}))\ ,
\end{equation}
where $\Delta_{Y_4} , *_{Y_4}$ is the Laplacian and the Hodge-star evaluated in the Calabi-Yau metric 
$g_{a \bar b}$. The last term in \eqref{Laplaceeq} needs to be included if the background contains M2-branes 
which fill the non-compact space-time $\bbR^{(2,1)}$ and are pointlike in $Y_4$.
There are further constraints by supersymmetry and the equations of motion on the background flux $\cG_4$.
It can be shown that $\cG_4$ 
has to be selfdual and primitive,
\beq \label{self-dualG}
 *_{Y_4} \cG_4 = \cG_4\,, \qquad J \wedge \cG_4 = 0\,,
\eeq
where $J$ is the K\"ahler form on the fourfold $Y_4$.
We will have to say more about the flux $\cG_4$ and its interpretation in the
Type IIB picture in section \ref{sec:CalcCorrections}.

Let us stress that for compact geometries the Laplace equation \eqref{Laplaceeq} 
implies a non-trivial consistency condition when integrated over $Y_4$. 
This is the famous M2-brane tadpole condition 
\beq \label{tadpole}
  \frac{\chi(Y_4)}{ 24} = \frac12 \int_{Y_4} \cG_4 \wedge \cG_4 + N_{\rm M2}\ ,
\eeq
where $\chi(Y_4)$ is the Euler number of $Y_4$, and $N_{\rm M2}$ is 
the number of space-time filling M2-branes. The condition \eqref{tadpole} 
together with \eqref{self-dualG} implies that in a compact setting the 
corrections due to $X_8$ leading to $\chi(Y_4)$ in \eqref{tadpole} are crucial 
to find supersymmetric vacua with $\cG_4$ flux. 
However, in our local considerations we will focus mainly 
on the flux contribution in \eqref{Laplaceeq}, and leave 
the inclusion of the curvature corrections to future work.

%%%%%%%%%%%%%%%%%%%%%%%%%%%%%%%%%%%%%%%%%%%%%%%%%%%%%%%%%%%%%%%%%%%%%%%%%%%%%%%
\subsection{Kaluza-Klein-monopoles: $\text{TN}_{k}$-spaces in M-theory} \label{BasicsTNk}
%%%%%%%%%%%%%%%%%%%%%%%%%%%%%%%%%%%%%%%%%%%%%%%%%%%%%%%%%%%%%%%%%%%%%%%%%%%%%%%

So far we have introduced the background geometry including a warped $Y_4$, and we recalled
that the warp factor is sourced by internal fluxes $\cG_4$. As a next step we like to 
identify local geometries in $Y_4$ which would 
correspond to D6-branes at weak coupling. Note that 
our geometries $Y_4$ will be elliptic fibrations 
in which such a weak coupling limit can be performed. The
D6-branes are located at the points where the elliptic fibration pinches. 
In particular, a D6-brane will wrap the divisors $S_{\rm b}$ in the 
base $B_3$ if the elliptic fiber pinches over this divisor. 
Clearly it is very hard to evaluate the warp factor equation \eqref{Laplaceeq}
for the full geometry $Y_4$. To proceed we therefore will focus on 
a local model denoted as $\mathcal{Y}_4$ which arises in a patch of $Y_4$ near $S_{\rm b}$. 

Before considering the periodic case with an additional circle 
let us first recall some classical facts about the origin of D6-branes 
in M-theory. The D6-brane is realized in M-theory
as a Kaluza-Klein monopole that is a  
solution to eleven-dimensional supergravity \cite{Townsend:1995kk}.
Roughly speaking, this monopole solution is  
an asymptotically locally flat circle fibration\footnote{The geometry approaches an $S^1$-bundle 
over $S^2\times \mathbb{R}$ at infinity in $\mathbb{R}^3$.} over $\mathbb{R}^3$ with 
degeneration loci at a point in $\mathbb{R}^3$. The asymptotic circumference 
of the circle fibration will be denoted by $r_{\rm A}$, and corresponds to the 
Type IIA string coupling 
\beq \label{gsIIA}
   g_s^{\rm IIA} = \frac{r_{\rm A}}{2\pi} \ .
\eeq
 In the weak coupling limit $r_{\rm A} \rightarrow 0$ the M-theory 
 setup reduces to the Type IIA string with a 
 D6-brane located at the point where the monopole circle pinches. 

We will directly consider the case of multiple Kaluza-Klein monopoles 
since we will need to consider periodic arrays later on. The solution 
with $k$ Kaluza-Klein monopoles will be denoted by $TN_{k}$. 
The metric of $TN_{k}$ is given by 
\beq \label{metric_TNk}
   ds^{2}_{TN_{k}} = \frac{1}{V} (dt + U)^2 + V d\vec r^{\, 2}\ ,
\eeq
where $t\sim t+r_{\rm A}$ is a periodic coordinate on a circle $S^1$ of circumference $r_{\rm A}=4 \pi m$
with $m$ 
being the mass of the Taub-NUT solution. The flat part of $TN_{k}$ is $\bbR^3$ with coordinates 
$\vec r = (x,y,z)$. The one-form $U$ on $\bbR^3$ is the $S^1$ connection.
In this metric one has the functions
\beq \label{eq:VmultiMonopole}
V = 1 + \sum_{i=1}^k V_I, \qquad  U = \sum_{I=1}^k U_I,\qquad V_I
= \frac{m}{{|\vec r-\vec{r}_I |}}, \qquad  *_3 \, d U_I = -d V_I\,,
\eeq
where $\vec r_I$ denote the positions of the $k$ monopoles, and $*_3$ is the 
Hodge star in $\bbR^3$. We denote this space as $TN_k$. We see that the circle fibration degenerates at the $k$ 
points $\vec{r}_I$ in $\bbR^3$.
Note that one has to use two patches around each monopole in order to obtain 
a globally well-defined connection $U_I$. Furthermore, one has to have 
the same mass $m$ for all monopoles in order 
to get a smooth solution.
The multi-center solution $TN_k$ admits $k$ anti-selfdual two-forms locally defined by
\beq \label{eq:DefOmegai}
\Omega_I = d\eta_I = \frac{1}{4 \pi m} d \Big(\frac{V_I}{V}(dt 
+ U) -  U_I \Big) ,\qquad I = 1,\ldots,k\ .
\eeq 
It is straightforward although technically involved to check that 
\beq \label{intersect}
\int_{TN_{k}} \Omega_I \wedge \Omega_J= - \delta_{IJ}\,, 
\eeq
as was noted in \cite{Ruback:1986ag} and is shown in detail in appendix \ref{app:TNgeo}.

Let us comment on the topology of $TN_k$. 
One can introduce the following real two-dimensional subvarieties of $TN_k$ defined as
\beq \label{eq:S_idef}
S_i = \big\{\left. (t, \vec r) \, \right|\, \exists p \, \in [0,1] \, 
\text{ s. t. } \, \vec r = (1-p)\vec{r}_i \,+ \, p\vec{r}_{i+1} \big\} , \qquad  i = 1,\ldots, k-1\ .
\eeq
These subvarieties are indeed closed two-cycles by noting the degeneration 
of the $S^1$-fiber at the position of the monopoles which gives them the 
topology of a sphere $S^2=\mathbb{P}^1$. The generators $S_1,\ldots, S_{k-1}$ span 
the second homology of $TN_k$ that is thus given by $\mathbb{Z}^{k-1}$.
Furthermore these surfaces intersect each other as the negative Cartan matrix $C_{ij}$ 
of $A_{k-1}$ which matches the fact that these geometries give $SU(k)$ 
gauge theories \cite{Sen:1997kz}. To see this one notices that $S_i$ and $S_{j}$, $i$ $\neq$ $j$, 
intersect each other exactly once if and only if $i$ = $j-1$ but with 
reversed orientation. To find the self-intersection of $S_i$, deform 
the base curve generically, which intersects the old one precisely at 
$\vec{r}_i$ and at $\vec{r}_{i+1}$ this time with the same orientation 
resulting in the self-intersection two. If we add the cycle $S_0$ connecting
$\vec{r}_1$ and $\vec{r}_{k}$, that is minus the sum of the $S_i$, we obtain the Cartan
matrix of affine $A_{k-1}$. This is consistent with the fact, that $TN_{k}$ is for
generic moduli the resolution of an $A_{k-1}$-singularity\footnote{Indeed if all 
monopoles approach each other the area of the $S_i$ vanishes and the 
space develops a $\mathbb{Z}_{k-1}$-singularity. To see this one expands 
the metric along the lines as was done for the case of the single 
monopole. The configuration which arises by squeezing all monopoles 
together corresponds to a monopole of charge $nm$ which equippes $\psi$ 
with a periodicity of $\frac{4\pi}{k}$ what shows 
the desired deficit angle.}. In summary $H_2(TN_k,\mathbb{Z})$ is isomorphic to the weight
lattice of $A_{k-1}$.

The Poincar\'e dual of $H_2(TN_k,\mathbb{Z})$ is given by $H^2_{\text{cpct}}(TN_k,\mathbb{Z})$,
the second cohomology with compact support. Hence it is isomorphic to $\mathbb{Z}^{k-1}$ and
its generators are given by  \cite{Bianchi:1996zj}
\beq \label{eq:omega_i-j}
    \hat \omega_i = \Omega_i - \Omega_{i+1}\ .
\eeq
These fulfill the following conditions, see appendix \ref{app:TNgeo},
\beq \label{Ptwoform}
\int_{TN_k} \hat \omega_i \wedge \hat \omega_j = - C_{ij}, \quad \int_{S_i} \hat \omega_j = - C_{ij} .
\eeq
This concludes our discussion of the space $TN_k$. It will be crucial in a next step to 
generalize these geometries to have infinitely many centers in order to describe periodic 
configurations. 

% We conclude by mentioning two further sensible second (co)homology groups on $TN_N$ discussed in 
% \cite{Witten:2009xu}. One is the cohomology $H^2(TN_N,\mathbb{Z})$ that can be 
% introduced by duality w.r.t.~a specific pairing with the cycles $S_i$ spanning $H_2(TN_N,\mathbb{Z})$. 
% This pairing is defined by first representing $H^2(TN_N,\mathbb{Z})$ by integers 
% $b_{i}$, that can be thought of as the duals to the points $\vec{r}_i$, $i=1,\ldots,N$, such that 
% $b_{i}\sim b_i +b$. Then we assign to the cycle $S_j$ the integer $b_{j+1}-b_j$. The  
% second group is the ``geometrical'' homology $H_2(TN_N,\mathbb{Z})$ spanned by the 
% $N$ non-compact cycles obtained by drawing parallel lines in the base $\mathbb{R}^3$ from 
% each point $\vec{r}_i$ to infinity, without hitting any other point $\vec{r}_j$, $j\neq i$. Including
% the circle fibration they become infinite ``cigars'' in $TN_N$ that can be shown to span the weight lattice
% of $U(N)$. Their Poincare dual forms are given by the anti-selfdual forms $\Omega_i$ in \eqref{eq:DefOmegai}.
% However, only $N-1$ of these cycles correspond to compact cycles in $TN_N$ and to massless
% modes in the effective theory of M-/F-theory on $TN_N$. \Dnote{Extension of this discussion by Thomas?}

%%%%%%%%%%%%%%%%%%%%%%%%%%%%%%%%%%%%%%%%%%%%%%%%%%%%%%%%%%%%%%%%%%%%%%%%%%%%%%%
\subsection{$S^1$-compactification of $\text{TN}_k$: $\text{TN}_{k}^\infty$-space in M-theory} 
\label{basics_TNinfty}
%%%%%%%%%%%%%%%%%%%%%%%%%%%%%%%%%%%%%%%%%%%%%%%%%%%%%%%%%%%%%%%%%%%%%%%%%%%%%%%

Our goal is to to eventually describe 7-branes F-theory and to derive their gauge coupling function. 
At weak Type IIB string coupling the corresponding D7-branes T-dualize to D6-branes localized on 
a circle, which we termed the B-circle. In order to describe this situation in M-theory we  
consider an infinite array of Kaluza-Klein monopoles separated 
by a distance $r_{\rm B}$ in the $z$-direction of $\mathbb{R}^3$ introduced in \eqref{metric_TNk}.
To effectively compactify this $z$-direction on a circle we mod out
the relation $z\sim z+r_{\rm B}$. This is analogous to the geometries considered 
in \cite{Ooguri:1996me,Blau:1997du,Eyras:1999at}. 
 
We first introduce the metric structure on the infinite array denoted by $TN^\infty_k$ in the following.
This space is obtained as follows. We first consider the special situation of $TN_k$ with centers located 
in the $(x,y)$-origin but separated along the $z$-coordinate in \eqref{metric_TNk} by a distance $z_I$.
This implies that we take the vectors $\vec r_I$ in \eqref{eq:VmultiMonopole} of the form 
\beq
   \vec r_I =  (0,0,z_I)\ , \qquad \quad 0 \le z_I < r_{\rm B}\ .
\eeq
Next we periodically extend this space to $TN^\infty_k$
in the $z$-direction with period $r_{\rm B}$.
The metric for such a configuration
still takes the form  
\beq \label{eq:TN_infty}
ds_{TN_k^\infty}^2 = \frac{1}{V}(dt + U)^2 + Vd\vec{r}^{\,2},
\eeq
where $V$ is a harmonic function on $\mathbb{R}^3$ except at the points $\vec{r}_I$ and  
$U$ a connection one-form,
\beq \label{PoissonV} 
 V = 1 + \sum_{I=1}^k V_I \,,\qquad  \quad U = \sum_{I=1}^k U_I\ . 
\eeq
Since we consider an infinite array $V_I, U_I$ are of the form 
\beq \label{per_Vi}
  V_I = \frac{r_{\rm A}}{4\pi}\sum_{\ell \in \mathbb{Z}} \frac{1}{\sqrt{\rho^2 + (z+\ell\, r_{\rm B} -z_I )^2}} 
  -  \frac{r_{\rm A}}{4\pi}\sum_{l \in \mathbb{Z}^*} \frac{1}{r_{\rm B} | \ell |} \ , \qquad \quad
  *_3\, d U_I = -dV_I\ . 
\eeq
where $r_{\rm A} = 4 \pi m$, $\bbZ^*= \mathbb{Z} \backslash \{0\}$, and $\rho$ = $\sqrt{x^2 + y^2}$. The first 
term in $V_I$ is just the potential of a periodic configuration of monopoles along the $z$-axis with spacing 
$r_{\rm B}$. The second term in $V_I$ is a regulator which ensures 
convergence and can be modified by any finite constant. 
This metric is also called the Ooguri-Vafa metric, that was initially constructed in the analysis of the 
hypermultiplet moduli space of Type II string theory \cite{Ooguri:1996me}. To see that the metric 
\eqref{eq:TN_infty} defined with $V$ and $U$ in \eqref{PoissonV} is smooth for finite and different 
$z_I \neq z_J$ for $I \neq J$ one notices that locally near the singularities of $V$ the space looks like that 
of one single Kaluza-Klein monopole which is known to be smooth. For our later discussion it will be crucial to 
introduce the rescaled coordinates
\beq \label{def-hatcoords}
  \hat t = \frac{t}{r_{\rm A}} \ , \qquad \hat z = \frac{z}{r_{\rm B}} \ , \qquad \hat z^I = 
  \frac{z^I}{r_{\rm B}}\ ,\qquad \hat \rho = \frac{\rho}{r_{\rm B}}\,.
\eeq
Note that in these coordinates one has the periodic identifications 
\beq
   \hat t = \hat t + 1\ , \qquad \hat z = \hat z + 1 \ , \qquad \hat z^I = \hat z^I + 1\ .
\eeq

To obtain a better understanding of the regularity and the physical meaning of the solution one has to perform a 
Poisson resummation of $V$ and $U$ \cite{Ooguri:1996me,Gaiotto:2008cd}. The details of the calculations are relegated 
to appendix \ref{app:TNchaingeo}. Finally we may then write 
\bea \label{correction}
V_I &=&  - \frac{r_{\rm A}}{2 \pi r_{\rm B}}\Big(\log\Big( \frac{\rho}{\Lambda r_{\rm B}}\Big) - \sum_{\ell \in 
\mathbb{Z}^*} K_0\Big(\frac{2\pi\rho}{r_{\rm B}} |\ell| \Big) \ e^{2\pi i \ell ( z - z_I) / r_{\rm B} }  \Big) 
\nn \\
&=&  - \frac{r_{\rm A}}{2 \pi r_{\rm B}} \Big( \log\Big(\frac{\hat \rho}{\Lambda}\Big) - 2 \sum_{\ell >0} 
         K_0(2\pi\hat \rho\, \ell ) \ \text{cos}( 2\pi  \ell (\hat z - \hat z_I)) \Big) \,,
\eea
where $\Lambda$ is a constant which can be chosen arbitrarily in the regularization of \eqref{per_Vi}.
\footnote{In appendix \ref{app:TNchaingeo} we have fixed  $1/\Lambda = \pi e^{2\gamma}$  
with $\gamma\approx0.577$ denoting the Euler-Mascheroni constant.} The function $ K_0(x)$ is the zeroth Bessel 
function of second kind.
Let us note that $V_I$ satisfies the Poisson equation 
\beq \label{eq:PoissonV_I}
    \Delta_{3} V_I = - \frac{r_{\rm A}}{r_{\rm B}\, \hat \rho}\ \delta(\hat z-\hat z_I) \delta(\hat \rho) 
    \delta(\varphi)\ ,
\eeq
where $\Delta_{3} = \frac{\partial^2}{\partial \hat \rho^2} + \frac{1}{\hat \rho} \frac{\partial}{\partial \hat 
\rho}+\frac{1}{\hat \rho^2} \frac{\partial^2}{\partial \varphi^2} + \frac{\partial^2}{\partial \hat z^2}$
is the Laplacian in cylinder coordinates. 
One can also perform a Poisson resummation for $U$, as we do in appendix \ref{app:TNchaingeo}, finding up to 
an ambiguity of an exact form
\bea \label{eq:Aresummed}
 U_I  &=&    \frac{r_{\rm A}}{4 \pi} \Big(-1- 2(\hat{z}-\hat{z}_I)   
      + 2i\hat{\rho} \sum_{\ell \in \mathbb{Z}^*} \text{sign}(\ell) K_1 \Big(2\pi\hat{\rho }  |\ell|\Big)  
      e^{2\pi i \ell (\hat{z} - \hat{z}_I) }   \Big)d\varphi \nn  \\
  &=& -\frac{r_{\rm A}}{4 \pi} \Big( 1+2(\hat{z}-\hat{z}_I)   + 4\hat{\rho} \sum_{\ell>0} K_1 (2\pi\hat \rho 
  \ell )\, \text{sin}\big({2\pi  \ell (\hat z - \hat z_I) }\big) \Big)d\varphi  \ , 
\eea
for $\hat{z}_I\leq \hat{z}<\hat{z}_I+1$, where $\varphi = \text{arctan}(y/x)$, and $K_1$ is the first Bessel 
function of second kind. In the first 
term in this expression we have included an integration constant $\hat{z}_I$ which 
arises when solving \eqref{per_Vi}. Note that this form is gauge equivalent to $U_I$ with leading term given by 
$U_I=\frac{r_{\rm A}}{2\pi}(\varphi_0+\varphi)d\hat{z}+\ldots$ by the gauge transformation by 
$d(\hat{z}\varphi)$. It will turn out below that it is important for the F-theory interpretation to define the 
full circle connection $U$ in this gauge reading
\beq \label{eq:Uwithdvarphi}
	U=\frac{k}{2\pi}r_{\rm A}(\varphi+\varphi_0)d\hat{z}-\frac{r_{\rm A}}{ 2\pi} \Big( 2  \hat{\rho} 
	\sum_{\ell>0} K_1 (2\pi\hat \rho \ell )\, \text{sin}\big({2\pi  \ell (\hat z - \hat z_I) }\big) \Big)d\varphi\,.
\eeq
Here we introduced an integration constant $\varphi_0$. As we will 
show next this choice of integration constant is required when matching the local geometry with an asymptotic  
elliptic fibration required in F-theory and equivalently for the identification of the three-dimensional RR-form 
$C_0\equiv k\varphi_0$.

For completeness we note that also the definition of the two-forms $\Omega_i$ can be extended to 
$TN_{k}^\infty$. They are given by 
\beq \label{def-Omegainf}
   \Omega^{\infty}_I = d\eta_I=\frac{1}{r_{\rm A}} d \Big(\frac{V_I}{V}(dt 
+ U) - U_I \Big) .
\eeq
As demonstrated in appendix \ref{app:TNchaingeo} these forms still satisfy 
\beq \label{TNin}
  \int_{TN_k^\infty} \Omega^\infty_I \wedge \Omega^\infty_J = - \delta_{IJ}\ , \qquad \quad *_4 \Omega^\infty_I 
  = - \Omega^\infty_I \ , 
\eeq
where the Hodge-star $*_4$ is in the $TN_k^\infty$ metric \eqref{eq:TN_infty}.
In addition, we introduce the generalization of the forms introduced in in \eqref{eq:omega_i-j} to the geometry 
$TN^\infty_k$,
\begin{equation}
	\omega_i^\infty=\Omega_i^\infty-\Omega_{i+1}^\infty\,.
\label{eq:omega^infty_i}
\end{equation}
As on the the Taub-NUT space $TN_k$ we expect them to generate the second cohomology with compact support 
$H^2_{\text{cpct}}(TN_k^\infty,\mathbb{Z})$ and to 
be dual to the connecting $\P^1$ between $z_i$ and $z_{i+1}$ of the resolved $A_{k-1}$ singularity. In 
particular, the intersections are given by the Cartan matrix $C_{ij}$ as in \eqref{Ptwoform}.

To close this section, let us now discuss the limit of large $\hat \rho$, which means 
that we are moving away from the centers of the monopoles.  
%Later on in 
%section \ref{sec:LeadingGaugeKin} we will relate this to the F-theory limit. 
In this limit one can expand 
\beq \label{largeK0}
K_0  (x) \sim \sqrt{\frac{\pi}{2x}}e^{-x}, \quad x \gg 1\,,
\eeq
so that the terms involving the Bessel functions in \eqref{correction} and \eqref{eq:Aresummed} 
are exponentially suppressed as $e^{- 2\pi \hat \rho |\ell| } \, \rightarrow\, 0$ for large $\hat \rho$.
Since the $z_I$ are the positions in the $z$-direction with period $r_{\rm B}$ this is equivalent to
smearing one Kaluza-Klein monopole along the $z$-direction in the base $\mathbb{R}^3$ to obtain a new 
isometrical direction.\footnote{In the picture of point particles in $\mathbb{R}^3$ this corresponds to a 
charged wire extended along the $z$-axis.}
One can then use this isometry to gauge away two components of the connection $ U$ keeping only the component 
$U_3$ in the $z$-direction. We therefore obtain the approximate potential and gauge connection
\beq \label{eq:leadingV+A}
V = 1 - \frac{k}{2 \pi} \frac{r_{\rm A}}{r_{\rm B}}\log\Big(\frac{\hat \rho}{\Lambda} \Big)\,, \quad 
  U  = \frac{k}{2 \pi } r_{\rm A} (\varphi + \varphi_0) d\hat z\,,
\eeq
up to leading order in $r_{\rm B}$.  Clearly, this means simply that we have dropped the exponentials in 
\eqref{correction} and \eqref{eq:Uwithdvarphi}. 
In the limit \eqref{eq:leadingV+A} we can rewrite the metric \eqref{eq:TN_infty} as 
\beq \label{metric}
ds_{TN_k^\infty}^2 \approx \frac{1}{V}\big((dt + U_z dz)^2 + V^2 dz^2 \big) + V(d\rho^2 + \rho d\varphi^2) \ ,
\eeq
where the coordinates have periods $(t,z) = (t+r_{\rm A},z+r_B)$.
In the next step we show that this is simply a two-torus bundle over the $(\hat \rho, \varphi)$-plane 
with metric 
\beq \label{torus-bundle}
   ds^2 = \frac{v_0}{ \I \tau} \big( (d \hat t + \R \tau \, d\hat z )^2 + (\I \tau)^2 d\hat z^2 \big) + ds_{\rm 
   base}^2\ ,
\eeq 
where $v_0$ is the volume of the two-torus fiber.
The rescaled coordinates $\hat t$ and $\hat z$ with integral periods were introduced already in 
\eqref{def-hatcoords}. Note that this torus structure is present due to the careful choice of boundary 
conditions, involving the constant $\varphi_0$ only, in the 
determination of \eqref{eq:leadingV+A}. Comparing \eqref{metric} and \eqref{torus-bundle} volume of the torus 
fiber is given by 
\beq 
   v_0 = r_{\rm A} r_{\rm B}\ .
\eeq 
The complex structure of the torus-fiber at a fixed point $u=\rho \, e^{i \varphi}$  in the $(\hat \rho, 
\varphi)$-plane, 
is given by 
\beq \label{eq:tauCalculated}
\tau(u) =  \frac{k}{2\pi} (\varphi_0+\varphi)+i \Big(\frac{r_{\rm B}}{ r_{\rm A}} -\frac{k}{2\pi} \, 
\log\Big(\frac{\hat 
\rho}{\Lambda}\Big)\Big) =\tau+\frac{k}{2\pi i}\log\Big(\frac{u}{\Lambda}\Big)\,.  
\eeq
Furthermore, the condition $dV$ $=$ $-*d U$ ensures that $\tau$ is a holomorphic function in $u$. Anticipating 
the discussion of F-theory in section \ref{limit}, we thus obtain 
precisely the expected monodromy of the the axio-dilaton in an F-theory with $k$ D7-branes at $u=0$. We identify 
the background value $\tau=C_0+ig_s^{-1}$ as
\beq \label{eq:identifyC0gs}
C_0=\frac{1}{2\pi}k\varphi_0\,,\qquad g_s=\frac{r_{\rm A}}{r_{\rm B}}\,.
\eeq
We also introduce the notation
\beq
\tau_I (u) =  \frac{1}{2\pi} (\varphi + \varphi_0) - i \frac{1}{2\pi}\, \log\Big(\frac{\hat \rho}
{\Lambda}\Big)\,.
\eeq
That the right-hand side of the equation carries no index is explained by the fact that we have neglected the 
subleading corrections.

For completeness and later reference we list the leading parts of the anti-selfdual two-form $\Omega$.
Inserting \eqref{eq:leadingV+A} and \eqref{eq:tauCalculated} into \eqref{def-Omegainf} we obtain
\beq \label{omega}
\eta_I^\infty = \frac{\I \tau_I}{\I \tau} (d\hat t + \R \tau d\hat z) -\R \tau_I d\hat z , \qquad 
\Omega^\infty_I = 
d\eta^\infty_I. 
\eeq
For the case of just one monopole we reproduce (in cohomology) the model discussed 
in \cite{Denef:2008wq} to describe a local 7-brane geometry.
\beq
\eta^\infty = \frac{r_{\rm B}}{r_{\rm A}} \frac{1}{\I \tau} (d\hat t + \R \tau d\hat z) , \qquad \Omega^\infty = 
d\eta^\infty.  
\eeq
As a next step one would have to construct the forms $\omega^\infty_i = \Omega^\infty_{i+1} - \Omega^\infty_i$
as in \eqref{eq:omega^infty_i}. However, having neglected the subleading corrections depending on the 
$z$-coordinate the forms $\omega^\infty_i $ would vanish identically for the forms \eqref{omega}. In other 
words, if we want to localize fluxes or gauge fields along the forms $\omega^\infty_i$ it will be crucial to 
include the non-trivial $z$-dependence in \eqref{correction}.

We conclude by interpreting the geometric meaning of the subleading exponential
sums in \eqref{correction} and \eqref{eq:Aresummed}. Approaching $\rho=0$ where the 
fiber torus degenerates, we note that the leading term of $V$ does not ``know'' about the position of the 
degeneration of the fibration of the A-circle 
\textit{on} the $z$-direction. The corresponding degenerated torus that arises from the 
leading term only, i.e.~the metric \eqref{metric}, merely looks like a very thin tire. 
However, the degenerated torus that arises from M-/F-theory should look like a torus that 
pinches at a point only, so that the pinched torus forms a $\P^1$. These two different pictures of the 
degeneration of the torus are 
called the ``differential geometric'' and the ``algebraic geometric'' degeneration in reference 
\cite{Aspinwall:1997eh}. Including now
the exponential corrections in $V$ and $U$, however, localizes the A-cyle degeneration and thus the torus 
degeneration at the point $z=0$ on the B-cycle, which reconciles the differential and
algebraic geometric pictures.

%%%%%%%%%%%%%%%%%%%%%%%%%%%%%%%%%%%%%%%%%%%%%%%%%%%%%%%%%%%%%%%%%%%%%%%%%%%%%%%
%%%%%%%%%%%%%%%%%%%%%%%%%%%%%%%%%%%%%%%%%%%%%%%%%%%%%%%%%%%%%%%%%%%%%%%%%%%%%%%
\section{7-brane gauge coupling functions in warped F-theory}
\label{sec:LeadingGaugeKin} 
%%%%%%%%%%%%%%%%%%%%%%%%%%%%%%%%%%%%%%%%%%%%%%%%%%%%%%%%%%%%%%%%%%%%%%%%%%%%%%%
%%%%%%%%%%%%%%%%%%%%%%%%%%%%%%%%%%%%%%%%%%%%%%%%%%%%%%%%%%%%%%%%%%%%%%%%%%%%%%%

In this section we turn to the computation of the gauge-coupling function of a 
stack of 7-branes in F-theory by using the dual M-theory. In order to 
do that we first recall some basics about F-theory on singular elliptically 
fibered Calabi-Yau fourfolds with an $A_{k-1}$ singularity 
along a divisor $S_{\rm b}$ in section \ref{limit}. This setup 
leads to an SU$(k)$ gauge theory in the effective four-dimensional 
theory, and has a weak coupling limit introduced in section \ref{sec:TypeIIBRev}.
In section \ref{limit} we also recall how F-theory can be viewed 
as a limit of M-theory. In section \ref{sec:LeadingPart} we use this map of F-theory 
to a dual three-dimensional M-theory compactification on a resolved 
Calabi-Yau fourfold to compute the leading gauge coupling function 
as in \cite{arXiv:1008.4133}. In order to include the corrections due to brane fluxes 
we perform a refined but local reduction in section \ref{sec:CalcCorrections}, and 
include a non-trivial warp factor and a back-reacted M-theory three-form as 
introduced in section \ref{sec:warpedreduction}. 
The resulting correction to the D7-brane gauge coupling 
can be matched with the weak coupling 
result of section \ref{sec:TypeIIBRev}.

\subsection{F-theory as a limit of M-theory} \label{limit}

To get started, let us recall some basic facts about a four-dimensional F-theory compactification on 
an elliptically fibered Calabi-Yau fourfold $Y_4$.
In general the elliptic fibration over a base $B_3$ is described by the 
Weierstrass form
\beq
y^2  = x^3 + f (\vec w)x^4 + g (\vec w) \ ,
\eeq
where $f (\vec w)$ and $g(\vec w)$ are sections of 
$K_{B_3}^{-4}$ respectively $K_{B_3}^{-6}$, and hence 
depend on the coordinates $\vec w$ of $B_3$. 
The modular parameter $\tau$ of the elliptic fiber
of $Y_4$ is only defined up to PSL(2, $\mathbb{Z}$) transformation and
thus most invariantly specified by 
\beq
\label{eq:j-function}
j(\tau(\vec w)) = \frac{4\cdot (24f )^3}{\Delta}\,, \qquad \Delta =27g^2 + 4f^3\ .
\eeq
where the $j$-function provides away from the singularities $\Delta = 0$ 
a biholomorphic map from the fundamental 
region to the complex plane. 
The fibration in particular implies, that $\tau$
is a section $\tau(\vec w)$ on the base $B_3$ and 
describes a varying coupling $\tau = C_0 + i e^{-\phi}$ of Type IIB. 
Clearly, near the singularities $\Delta = 0$ this simple picture 
breaks down and has to be replaced by a refined local treatment 
as we discussed above in section \ref{basics_TNinfty}.

The special subloci on $B_3$ where the discriminant $\Delta$ vanishes
indicate the presence of objects charged under $\tau$. These 
loci geometrically describe divisors in $B_3$ over which the elliptic 
fiber becomes singular. In Type IIB string theory these divisors are 
wrapped by $(p,q)$7-branes. The particular type of fiber 
degeneration leads to different monodromies of $\tau$
around the singular divisors that encode the type of $(p,q)$7-branes and 
the gauge groups on these branes. 
As an example we consider a singular $Y_4$ with an $A_{k-1}$ singularity in the elliptic fiber 
over a divisor $S_{\rm b} \subset B_3$ which describes a
stack of $k$ D7-branes on  $S_{\rm b}$. In other words 
we consider the split of the class $[\Delta]$ of the discriminant as 
\beq
   [\Delta] = k [S_{\rm b}] + [\Delta']\ ,
\eeq
where $[\Delta']$ is the residual part of $\Delta$ wrapped by a single 
complicated 7-brane. While $\Delta'$ might intersect $S_{\rm b}$ 
the new physics at these intersections will not be of crucial importance to the discussion 
of this work. We will mainly focus on a local model near $S_{\rm b}$
and concentrate on the back-reaction of the flux on the geometry. 
In this local model we introduce a local complex coordinate $u$  
such that $S_{\rm b}$ is given by $u=0$.
In the vicinity of $S_{\rm b}$
we have the local behaviour
\beq
j(\tau(\vec w)) = a\frac{1}{u^k} + b \quad \Rightarrow \quad \tau(\vec w) = 
\begin{cases} j^{-1}(b) & \text{far way from the D7-branes}\\ -i\frac{k}{2\pi} \log(u) & 
\text{near the D7-branes} \end{cases}
\eeq
where we have used that $j(\tau)$ $\sim$ $e^{-2\pi i \tau}$ for large Im($\tau$). 
This is precisely the naively expected dilaton in the neighborhood of a D7-brane
in perturbative Type IIB theory.

Before turning to the discussion of the formulation of F-theory via M-theory, 
let us make contact with the presentation of section \ref{sec:TypeIIBRev}. In this section 
we have considered the weakly coupled limit of F-theory \cite{Sen:1996vd,Sen:1997gv}.  
In this very special case the axio-dilaton $\tau(u)$ is constant almost everywhere on 
$B_3$ and chosen to have $\I \tau \gg 1$ corresponding to 
a small Type IIB string coupling $g_{\rm IIB}$. The fundamental objects are in this limit 
D7-branes and O7-planes. 

To study F-theory compactifications away from the weak coupling limit is in 
general a hard task. The complication arises due to the fact that there is no fundamental 
twelve-dimensional effective action for F-theory which could be used at low energies. 
To nevertheless investigate general F-theory configurations one has to take a detour 
via M-theory and a three-dimensional compactification. One starts with a 
compactification of 
M-theory on the singular elliptically fibered Calabi-Yau fourfold $Y_4$. Due 
to the singularities of $Y_4$ there are massless M2-brane states which are 
massless and generate a non-trivial gauge theory and spectrum 
in the effective three-dimensional theory. For our setups with an 
$A_{k-1}$ singularity over $S_{\rm b}$ one finds a three-dimensional 
non-Abelian gauge theory with gauge group $G=SU(k)$. The F-theory 
limit is performed by shrinking the volume of the elliptic fiber of $Y_4$. Since 
the A-circle shrinks, this yields a Type IIA compactification on a small B-circle over 
$B_3$. After T-duality along the B-circle this will yield a Type IIB string compactification 
on $B_3$ times the T-dual B-circle. In the F-theory limit this growing extra circle 
yields an additional non-compact direction and hence the effective theory will 
be four-dimensional. In our discussion it will be crucial to include the warp factor 
in the general M-theory solution \eqref{warped-background} when performing this duality. 

To actually derive the couplings of this theory one can resolve $Y_4$ 
to obtain a smooth geometry $\hat{Y}_4$, on which one can Kaluza-Klein reduce 
eleven-dimensional supergravity \eqref{Mtheoryaction}. Geometrically this 
yields $k-1$ new exceptional divisors $D_i$ in $\hat{Y}_4$ resolving 
the $A_{k-1}$ singularity over $S_{\rm b}$. We denote the 
Poincar\'e dual two-forms to $D_i$ by $\omega_i$. The 
Kaluza-Klein reduction of M-theory to three dimensions 
requires to expand the K\"ahler 
form $J$ of $Y_4$, as well as the M-theory three-form potential $C_3$
into harmonic modes. Explicitly, one has \footnote{Note that we restrict to 
Calabi-Yau fourfolds with $h^{2,1}(\hat{Y}_4)=0$, such that no extra scalars arise from $C_3$.}
\bea \label{JC3-expansion}
   \frac{J}{\cV} & = & R\, \omega_0 + L^\alpha \omega_\alpha + \xi^i \omega_i \\
   C_3 &=& A^0 \wedge \omega_0 + A^\alpha \wedge \omega_\alpha + A^i \wedge \omega_i\ ,  \nn
\eea
where $\cV$ is the volume of the Calabi-Yau fourfold $\hat{Y}_4$.
Here we have included the two-form $\omega_0$ Poincar\'e dual to the base $B_3$, 
and the two-forms $\omega_\alpha$ Poincar\'e dual to divisors $D_\alpha = \pi^{-1}(D_{\alpha}^{\rm b})$
inherited from divisors $D^{\rm b}_\alpha$ of the base. The coefficients 
$(R,L^\alpha, \xi^i)$, and $(A^\cA)=(A^0,A^\alpha, A^i)$, with $\cA$ $\in$ $\{\alpha, \, 0, i\}$, are real scalars and 
vectors in the 
three-dimensional effective theory. In the F-theory limit to four dimensions, the vector multiplet with bosonic 
components $(R,A^0)$ becomes part of the four-dimensional metric, and one 
identifies 
\beq
    R = r_{\rm B}^2 \ ,
\eeq
where $r_{\rm B}$ is circumference of the circle on which the T-duality to 
Type IIB is performed, and $\cV$ is the volume of $\hat{Y}_4$. 
The vector $A^0$ is the Kaluza-Klein vector in 
the four-dimensional metric as in \eqref{KKa}. The vector 
multiplets with bosonic components $(L^\alpha,A^\alpha)$ lift to complex scalars $T_\alpha$ in 
the F-theory limit, just as in appendix \ref{app:linMultis+reduction}. Finally, the vector 
multiplets with bosonic components $(\xi^i,A^i)$ lift to four-dimensional U$(1)$ vector multiplets 
gauging the Cartan generators $T_i$ of the four-dimensional SU$(k)$ gauge group as in section 
\ref{CYorientifolds}.

In order to proceed further in the discussion, let us recall 
the behavior of the fields in the F-theory lift. The latter is given 
by the vanishing of the fiber volume and the blow-down map from $\hat{Y}_4$ to 
$Y_4$. To make this more precise we introduce the following $\epsilon$-scaling \cite{arXiv:1008.4133}
\beq \label{eq:FtheoryLift}
r_{\rm B} \mapsto \epsilon\, r_{\rm B}\ , \qquad R \mapsto \epsilon^2 R\ , \qquad \zeta^i \equiv \frac{\xi^i}{R} 
\mapsto \epsilon^{2/3} \zeta^i \,, \qquad L^{\alpha} = 2L^{\alpha}_{\rm IIB}\, , 
\eeq
where the scalars $L^\alpha$ do not scale with $\epsilon$ but are identified with a factor two with the Type IIB 
variables $L^{\alpha}_{\rm IIB}$ 
used in appendix \ref{app:linMultis+reduction}. Note that the Type IIB string coupling is 
given by $g_s^{\rm IIB} = r_{\rm A}/r_{\rm B}$.
as can be inferred by using the T-duality rules applied to the Type IIA coupling \eqref{gsIIA}. 
Since $g_s^{\rm IIB}$ should not scale in the F-theory limit, we find that also $r_{\rm A} \mapsto \epsilon\, 
r_{\rm A}$.
We note in addition that this identification of the string coupling perfectly agrees with 
\eqref{eq:identifyC0gs} from M-
theory on $TN_k^\infty$.

We can thus give a diagrammatic summary of the limit we 
will consider. 
Recalling all identifications from M-theory on $TN^k_\infty$ in section \ref{M-theory_geom},
\beq \label{eq:defcouplings}
  v^0 = r_{\rm A} r_{\rm B},\qquad  g_s^{\rm IIA}= \frac{r_{\rm A}}{2\pi}\ , \qquad g_{s}^{\rm IIB}=\frac{r_{\rm 
  A}}{r_{\rm B}}\ ,  
\eeq
we consider the following limits:
\begin{equation}
\label{eq:diagram}
	\xymatrixcolsep{5Em}\xymatrixrowsep{4Em}\xymatrix %@C=1cm 
	{
  \parbox{4cm}{\centering M-theory on $TN_\infty$ \\ $r_{\rm A}$, $r_{\rm B}$ finite}\ar@{->}[r]^{\text{F-limit}}_{v^0 
  \rightarrow 0}\ar@{->}[d]^{g_s^{\rm IIA}\rightarrow 0} &  \parbox{5cm}{\centering 10d F-theory \\  
  $g^{\rm IIB}_s$ finite}\ar@{->}[d]^{g^{\rm IIB}_s\rightarrow 0}\\ 
  \parbox{4cm}{\centering Type IIA in 9d\\
  $r_{\rm A}$ finite, $g_s^{\rm A}\sim 0$}\ar@{->}[r]^{\text{F-limit}}_{v^0\rightarrow 0} &\parbox{5cm}
  {\centering weakly coupled 10d  IIB\\  $g_s^{\rm IIB}\sim 0$\,.}}
\end{equation}
Understanding the geometry and the physics of the four corners of this diagram is
essential for the calculations of the corrections to the gauge kinetic function in section
\ref{sec:CalcCorrections}. 

It is important for us to also follow the space $TN_k^\infty$ through the M-theory to F-theory lift. 
In fact, since the space $TN_k$ corresponds in Type IIA to $k$ parallel D6-branes, the space 
$TN_k^\infty$ yields an infinite array of periodically repeating parallel D6-branes. The 
periodic coordinate in section \ref{basics_TNinfty} was $z = z+r_{\rm B}$, which we normalized to 
have integer periods by setting $\hat z=z/r_{\rm B}$. In the $z$-direction the monopoles 
are separated by distances $z_{i+1} - z_{i}$, where $z_I$ are the locations of the $k$ monopoles. 
Without loss of generality we will take in the following $z_1 = 0$, setting the location of the first 
monopole to be the origin.
We identify the blow-up modes $\xi^i$ in \eqref{JC3-expansion} with the normalized differences as we will 
later justify in section \ref{sec:corrImaginarypart} as  
\beq
     \xi^i = r_{\rm B}(z_{i+1} - z_{i})\ . 
\eeq
In the F-theory limit $\epsilon \rightarrow 0$ the vanishing of the $\xi^i$ requires 
to also moving the centers on top of each other by sending $z_{i+1} \rightarrow z_{i}$, i.e.~one has
to send $z_I \rightarrow 0$.

%%%%%%%%%%%%%%%%%%%%%%%%%%%%%%%%%%%%%%%%%%%%%%%%%%%%%%%%%%%%%%%%%%%%%%%%%%%%%%%
\subsection{Leading 7-brane gauge coupling functions}
\label{sec:LeadingPart}
%%%%%%%%%%%%%%%%%%%%%%%%%%%%%%%%%%%%%%%%%%%%%%%%%%%%%%%%%%%%%%%%%%%%%%%%%%%%%%%

In this section we recall how the classical volume parts of 
the 7-brane gauge coupling function can be derived in F-theory via M-theory. 
This derivation only involves topological methods and can therefore be treated 
in a rigorous global picture of a compact Calabi-Yau fourfold $Y_4$. 
We return to a local analyis when deriving the corrections to the 
gauge-coupling function in section \ref{sec:CalcCorrections}.

Let us note that the field strength of $C_3$ given in \eqref{JC3-expansion}
is given by 
\beq \label{G4exp}
   G_4 = F^\cA \wedge \omega_{\cA} = F^{0} \wedge \omega_0 + F^{\alpha} \wedge \omega_{\alpha} + F^i \wedge 
   \omega_i\ .
\eeq
In this expression $F^\cA = d A^\cA$ are the field strengths of the three-dimensional 
$U(1)$ gauge fields.
%The last term in \eqref{G4exp} is the background flux $\cG_4$ 
%entirely supported on $\hat{Y}_4$. 
The three-dimensional effective action is computed by inserting the 
expansion \eqref{G4exp} into the eleven-dimensional supergravity 
action \eqref{Mtheoryaction}. Since we are interested 
in the leading flux-independent gauge coupling function 
we assume here that the metric is not warped by demanding  
that the warp factor $e^{3A/2}$ in \eqref{warped-background} is constant, and we 
set the background flux $\cG_4=0$.
Here we are interested in the reduction of the kinetic term of $G_4$, and 
derive 
\bea \label{eq:gaugeKinLeading}
S^{(11)}_{\text{kin}} =\frac{\pi}{2}\int G_4 \wedge * G_4 \cong 2 \pi  \int_{\mathbb{R}^{(2,1)}} \cG_{\cA \cB}\, 
F^\cA\wedge * F^\cB \eea
where in the second equality we have performed a Weyl rescaling of the three-dimensional metric $g^{(3)} 
\rightarrow \cV^2 g^{(3)}$ 
in order to bring the action into the Einstein frame, and 
introduced the metric 
\beq \label{cGAB}
   \cG_{\cA \cB} = \frac{\cV}{4} \int_{\hat{Y}_4} \omega_\cA \wedge * \omega_{\cB}\ , \qquad \omega_{\cA} = 
   (\omega_0,\omega_i,\omega_{\alpha})
\eeq
In the following we compute the metric $\cG_{\cA \cB}$ explicitly 
and discuss the matching with \eqref{eq:C2actionred} in 
order to read off the gauge-coupling function.

In order to compute the metric $\cG_{\cA \cB}$ 
explicitly we need some information about the intersections 
of the various forms $\omega_\cA$. We define $\cK_{\cA \cB \cC \cD} = \int_{\hat{Y}_4} \omega_\cA \wedge 
\omega_{\cB}\wedge \omega_\cC \wedge \omega_{\cD}$.
Due to the elliptic fibration structure one has $\cK_{\alpha\beta\gamma\delta}=0$. In addition we have $\omega_i 
\wedge \omega_0 =0$ in cohomology.
We will need the following non-vanishing intersections\footnote{We note the additional 
factor of $\frac{1}{2}$ in \eqref{eq:intsX4} in the definition of the intersection numbers $\mathcal{K}$
that was included in \cite{Grimm:2011tb} to identify with the intersections of the orientifold geometry 
$B_3=Z_3/\mathcal{O}$ in the upstairs-picture.}
\beq \label{eq:intsX4}
 \cK_{0 \alpha \beta \gamma} \equiv\tfrac{1}{2} \cK_{\alpha \beta \gamma} \,,\qquad \quad
 \cK_{ij \alpha \beta}=  - \tfrac12   C^\gamma_{ij}\, \cK_{\alpha \beta \gamma}\ , \qquad C^\gamma_{ij} \equiv 
 C_{ij} C^\gamma\ ,
\eeq
where $C_{ij}$ denotes the Cartan matrix of $G$ as above in section \ref{sec:TypeIIBRev}.
We recall that in the M-theory reduction the complex coordinates are given by 
\bea
   T_\alpha &=& \tfrac16\int_{D_\alpha} J \wedge J \wedge J + i \int_{D_\alpha} C_6 \\
            &=& \tfrac{1}{4} \cV^3  \cK_{\alpha \beta \gamma}  \big(L^\beta L^\gamma R-C^\gamma_{ij} L^\beta  
            \xi^i \xi^j ) + i \rho_\alpha + \ldots \ ,
\eea
where we have used \eqref{JC3-expansion}.
Using the intersections \eqref{eq:intsX4} one evaluates \footnote{Strictly speaking, a precise match requires a 
coodinate redefinition of the $L^\alpha$ 
with a term proportional to $C^\alpha_{ij}\xi^i \xi^j /R$ as in reference \cite{Bonetti:2011mw}. We will omit 
this here for simplicity. The factors can be fixed by matching the terms which are unaffected by this shift.}
\bea
   \cG_{ij} &=& \frac{C^\alpha_{ij}}{4 R} \R T_\alpha + \cG_{\alpha \beta } \frac{C^\alpha_{i k} C^\beta_{j l} 
   \xi^k \xi^l}{R^2} + \ldots\ ,\\
   \cG_{i\alpha} &=& -\frac{\cG_{\alpha \beta}}{R} C^\beta_{ij} \xi^j + \ldots   \ ,
\eea
where the dots indicate terms which are of higher power in $R$.
Inserting these expressions into \eqref{eq:gaugeKinLeading} we 
find the action
\beq
  S^{(3)}_{\rm kin} = - 2 \pi \int \cG_{\alpha \beta} (F^\alpha - R^{-1} \xi^i C^\alpha_{ij} F^j)\wedge * (F^\beta - R^{-1} \xi^i C^\beta_{ij} F^j)
                      + \tfrac{1}{4 R} C^\alpha_{ij} \R T_\alpha F^i \wedge * F^j\ . 
\eeq
Comparing this action with \eqref{eq:C2actionred} we infer that the leading gauge coupling function is 
simply given by 
\beq \label{fijM}
   f_{ij} = \tfrac12 C^\alpha_{ij}  T_\alpha = \tfrac14 C^\alpha_{ij} T_\alpha^{0\, {\rm IIB}}  \ ,
\eeq
where we recall from \eqref{eq:FtheoryLift} that we have to identify $T_{\alpha}$ = $\frac{1}{2}T^{0\, {\rm 
IIB}}_\alpha$. 
Note that this expression agrees with the weak coupling result \eqref{eq:fIJD7} if 
we drop the correction term $Q_\alpha$ containing the flux. It will be the task of the final subsection to also 
reproduce this correction. 

Let us conclude this section by noting that the expression \eqref{fijM} can 
also be directly infered from an M-theory kinetic potential $\tilde K$. It was 
shown in \cite{Grimm:2011tb} that for an elliptic fibration it takes the form 
\beq \label{tildeKM} 
  \tilde K^{\rm M} = \log \Big(\tfrac{1}{12} R L^{\alpha} L^{\beta} L^{\gamma} \cK_{\alpha \beta \gamma} 
-\tfrac{1}{8}\xi^i \xi^j C^\alpha_{ij}L^{\beta}L^{\gamma} \cK_{\alpha \beta \gamma} 
+ \ldots \Big) .
\eeq
and can be obtained from a K\"ahler potential given by $K^{\rm M} = - 3 \log \cV$ 
via a Legendre transform.
If one Taylor expands \eqref{tildeKM} 
around the F-theory point in moduli space with small $\xi^i$ one finds
\beq \label{Taylor}
\tilde K^{\rm M} =  \log \big(\tfrac{1}{12} L^{\alpha} L^{\beta} L^{\gamma}
K_{\alpha \beta \gamma}\big) + \log(R) - \frac{C_{ij}^\alpha \cK_{\alpha \beta \gamma}L^\beta L^\gamma}{\frac{1}
{3}\cK_{\alpha \beta \gamma}R L^\alpha L^\beta L^\gamma} \xi^i\xi^j .
\eeq
with $\cK_{\alpha\beta\gamma}$ the intersection numbers \eqref{eq:intsX4}.
Comparing this form with the general expression \eqref{Kahler3} of a 
three-dimensional kinetic potential one confirms the identification \eqref{fijM} 
of the classical gauge coupling function.

%%%%%%%%%%%%%%%%%%%%%%%%%%%%%%%%%%%%%%%%%%%%%%%%%%%%%%%%%%%%%%%%%%%%%%%%%%%%%%%
%%%%%%%%%%%%%%%%%%%%%%%%%%%%%%%%%%%%%%%%%%%%%%%%%%%%%%%%%%%%%%%%%%%%%%%%%%%%%%%
\subsection{On dimensional reduction with fluxes and warp factor}
\label{sec:warpedreduction}
%%%%%%%%%%%%%%%%%%%%%%%%%%%%%%%%%%%%%%%%%%%%%%%%%%%%%%%%%%%%%%%%%%%%%%%%%%%%%%%

In this subsection we discuss the dimensional reduction of M-theory with a 
warp factor and background four-form fluxes $\cG_4$. Our main focus will 
be on the modifications arising in the reduction of the M-theory three-form. 
Our results will extend the discussion in \cite{Dasgupta:1999ss}. 

Let us now perform the reduction including the warp factor. 
For simplicity we 
will not include higher curvature corrections and mobile M2-branes in 
the supergravity action \eqref{Mtheoryaction}. 
We will focus on the terms involving $G_4$ only, i.e.~the kinetic 
terms and the Chern-Simons term.
For the M-theory three-form $C_3$ itself we make the reduction Ansatz 
\beq \label{C3_expand}
  C_3 = A^\cA \wedge \tilde \omega_\cA + \beta(M^\Sigma) \ ,
\eeq
where $\tilde \omega_\cA$ are two-forms and $\beta$ is a three-form 
on $\hat Y_4$. The fluctuations are parameterized by three-dimensional 
vectors $A^\cA$ and scalars $M^\Sigma$, which change the geometry of $\hat Y_4$.
To restrict to the case of massless vectors $A^\cA$ we demand in the following
\beq
  d \tilde \omega_\cA = 0\ .
\eeq
We introduce the three-forms
\beq
   \beta_\Sigma = \frac{\partial \beta}{\partial M^\Sigma}
\eeq
The three-form $\beta$ is only patchwise defined, since we demand 
that in cohomology  $d_8 \beta$ encodes the topologically non-trivial background flux $\cG_4$.
This yields the field strength  
\beq \label{G4_expand}
  G_4 = F^\cA \wedge  \tilde \omega_\cA  + dM^\Sigma \wedge \beta_{\Sigma} + \cG_4\ .
\eeq

On next inserts the expressions \eqref{C3_expand} and \eqref{G4_expand} in the 11d
supergravity action 
\beq
 S^{(11)}_{G_4} = 2 \pi \int \frac{1}{4} G_4 \wedge * G_4 + \frac{1}{12} C_3 \wedge G_4 \wedge G_4 \ ,
\eeq 
using the warped metric \eqref{warped-background}. 
In order to bring the Einstein-Hilbert term into the standard 3d from 
one has to perform a Weyl rescaling with the warped volume 
\beq
   \cV_w = \int_{\hat Y_4} e^{3A/2} J\wedge J\wedge J\wedge J\ .
\eeq
As a result one finds the 3d action \footnote{Note that the reduction of the Chern-Simons term is complicated by the 
fact that the M-theory potential $C_3$ appears without derivatives.  We suppress terms of the form  
$\beta\wedge \partial_{M^{\Lambda}}\beta_\Sigma$ which are manifestly not gauge invariant. Terms of this type appear 
in Chern-Simons couplings for D-branes and it would be interesting to interpret them. These terms can be computed 
explicitly in our example and vanish for the derivatives w.r.t. the $M^\Sigma$ we study.}
\bea \label{red_warped_action}
 S^{(3)}_{G_4} &=& 2 \pi \int_{\cM_3} \cG_{\cA \cB}^w F^\cA \wedge * F^\cB +  d_{\Sigma \Lambda}^w d M^\Sigma \wedge * dM^\Lambda + V_w\, *1  \\
               &&\phantom{ 2 \pi \int_{\cM_3}} + \tfrac14\Theta_{\cA \cB} A^\cA \wedge F^\cB + d_{\cA \Sigma \Lambda}( M^\Sigma dM^\Lambda \wedge F^\cA) \ . \nn
\eea 
We discuss the various terms appearing in this action in turn. Firstly, there is the kinetic term for the 
vectors $A^\cA$ with coupling  
\beq \label{def-cGw}
   \cG_{\cA \cB}^w = \frac{\cV_w}{4} \int_{\hat Y_4} e^{3A/2} \tilde  \omega_\cA \wedge * \tilde \omega_\cB\ .
\eeq
Note that in contrast to \eqref{cGAB} a warp factor appears in the integral. By solving the 
warp factor equation \eqref{Laplaceeq} we will later show that this induces a flux correction to the gauge coupling function.  
The term involving $d_{\Sigma \Lambda}^w$ is a correction to the kinetic term of the scalars $M^\Sigma$. Its explicit form reads
\beq
d_{\Sigma \Lambda}^w = \frac{1}{\cV_w}\int_{\hat Y_4} \beta_\Sigma \wedge * \beta_\Lambda  \, ,
\eeq
where we have performed a Weyl rescaling and the Hodge star refers to the unwarped metric, i.e. this term 
happens to be independent of the warp factor.
Moreover, there is the well-known 3d potential $V_w$ introduced by the 
background flux $\cG_4$. 

The terms in the second line of \eqref{red_warped_action} arise from the reduction of the 11d Chern-Simons coupling. 
The term proportional to $\Theta_{\cA \cB}$ is a three-dimensional Chern-Simons term with constant coefficient 
$\Theta_{\cA \cB} = \int_{\hat Y_4} \cG_4 \wedge \tilde \omega_\cA \wedge \tilde \omega_\cB$. Depending on the 
index structure this Chern-Simons 
term either induces a gauging for non-trivial $\Theta_{i\alpha}$ in the dual 4d F-theory 
compactification \cite{arXiv:1008.4133,Grimm:2011sk,Grimm:2011tb}, or for $\Theta_{ij}$ generated at one loop by the 
four-dimensional chiral matter \cite{Grimm:2011fx}. Finally, the last term in \eqref{red_warped_action} contains the 
coupling 
\beq
   d_{\cA \Sigma \Lambda} = -\frac{1}{4}\int_{\hat Y_4} \tilde \omega_\cA \wedge \beta_\Sigma \wedge 
   \beta_\Lambda\ .
\eeq
We will later show that coupling induces a flux correction to the imaginary part of the F-theory 
gauge coupling function.

%%%%%%%%%%%%%%%%%%%%%%%%%%%%%%%%%%%%%%%%%%%%%%%%%%%%%%%%%%%%%%%%%%%%%%%%%%%%%%%
%%%%%%%%%%%%%%%%%%%%%%%%%%%%%%%%%%%%%%%%%%%%%%%%%%%%%%%%%%%%%%%%%%%%%%%%%%%%%%%
\subsection{Calculation of corrections to the gauge kinetic function}
\label{sec:CalcCorrections}
%%%%%%%%%%%%%%%%%%%%%%%%%%%%%%%%%%%%%

Finally we are well equipped in order to derive the correction to the 
gauge kinetic function induced by a non-trivial background flux $\cG_4$. 
We will show that these corrections match in the weak coupling limit the well-known 
corrections to the gauge kinetic function due to D7-brane flux.

The basic idea to compute the corrections to the real part of the gauge coupling function 
\eqref{fijM} is to derive the gravitational back-reaction of the fluxes on the warp factor 
in M-theory via \eqref{Laplaceeq}. This computation requires an explicit knowledge of the 
metric on the M-/F-theory fourfold $\hat{Y}_4$. We describe the elliptic fourfold 
$\hat{Y}_4\rightarrow Y_4$ with a resolved SU$(k)$ singularity in the elliptic 
fibration locally in the vicinity of the resolved singularity by the local geometry constructed 
in section \ref{basics_TNinfty},
\beq
\mathcal{Y}_4 = S_{\rm b} \times TN^\infty_k\,.
\eeq
Here $S_{\rm b}$ is that divisor in the base $B_3$ of the elliptic fibration\footnote{We 
focus here on SU$(k)$-singularities only  in co-dimension $1$ in $B_3$, i.e.~$S_{\rm b}$ is the full internal 
world-volume of the wrapped branes in a D-brane picture.} $Y_4$ with 
the SU$(k)$-fibre singularity. $TN^\infty_k$ is the periodic chain of multi-center Taub-NUT spaces 
with metric \eqref{eq:TN_infty}, that locally describes the normal space in $\hat{Y}_4$ to the 
resolved singularity over $S_{\rm b}$. As discussed in section \ref{basics_TNinfty}, the metric on $TN_k^\infty$ 
is known and governed by the function $V=1+\sum_{I=1}^k V_I$ and the gauge 
connection $U$ of \eqref{correction} respectively \eqref{eq:Aresummed}. 

In a brane picture in Type IIA and IIB or F-theory, the compactification of M-theory on 
$\mathcal{Y}_4$ describes the Coulomb branch with U(1$)^k$ gauge symmetry of the 3-dimensional 
gauge theory from $k$ parallel spacetime-filling 6-branes or T-dual, fluxed $k$ 7-branes 
wrapping $S_{\rm b}\times \mathcal{M}_3$ respectively $S_{\rm b}\times S^1\times\mathcal{M}_3$\footnote{Note 
that the flux on the 7-brane is T-dual to the separation of 6-branes on $S^1$, i.e.~has one leg 
on $\mathcal{M}_3$ and one leg on $S^1$. It breaks $U(k)\rightarrow U(1)^k$ and is not to be 
confused with the fluxes $\mathcal{F}^i$ introduced next in \eqref{eq:G4fluxOnTN}.}, where $S^1$ 
denotes the circle in the basis of Taub-NUT $TN_k^\infty$, $\mathbb{R}^2\times S^1$.
In this picture we also introduce the localized $\cG_4$-flux in M-theory. This flux is 
identified with two-form flux $\hat{\cF}^I$ of the $I$-th 6-brane on $S_{\rm b}$ or its T-dual 
7-brane that is valued in the U(1$)$ gauge group of the corresponding brane. It can be embedded 
into the Cartan subalgebra of the enhanced gauge group U(1$)\times $SU$(k)$ by defining new fluxes $\cF^0$ and $\cF^i$, $i=1,\ldots, k-1$, as  
\beq \label{eq:fluxOnTN}
 \hat \cF^{m} = \cF^0+\cF^{m} - \cF^{m-1}\,,\qquad \hat \cF^{1} = \cF^0+\cF^{1}\,,\qquad \hat \cF^{k} = \cF^0- \cF^{k-1}\,,
\eeq
where $m=2,\ldots, k-1$.
The flux on $\mathcal{Y}_4$ is thus of the form 
\beq \label{eq:G4fluxOnTN}
  \cG_4 = \hat \cF^I \wedge \Omega^{\infty}_I = \cF^i \wedge \omega_i^\infty+\cF^0\wedge\sum_J 
  \Omega_J^\infty\,, 
\eeq
where the second equality can be checked easily using \eqref{eq:fluxOnTN} and where $I=1,
\ldots,k$ and  $i=1,\ldots,k-1$. Recall that $\omega^\infty_i = \Omega^\infty_{i} - 
\Omega^\infty_{i+1}$ are two-forms on $TN_k^\infty$ which 
have been introduced already in \eqref{eq:omega^infty_i}, and satisfy $\int_{TN_k^\infty}
\omega^\infty_i \wedge \omega^\infty_j = -C_{ij}$.
Note that these forms should be identified with the blow-up forms $\omega_i$ appearing 
in \eqref{JC3-expansion} in the global embedding. Note that the two-form in the expansion
with $\cF^0$ is trivial in cohomology in $TN_k^\infty$, which matches the fact that the 
corresponding diagonal U$(1)$ in the enhancement to gauge group U$(k)$ is massive and integrated 
out in the effective theory.

%%%%%%%%%%%%%%%%%%%%%%%%%%%%%%%%%%%%%%%%%%
%%%%%%%%%%%%%%%%%%%%%%%%%%%%%%%%%%%%%%%%%%%%%%%%%%%%%%%%%%%%%%%%%%%%%%%%%%%%%%%

%%%%%%%%%%%%%%%%%%%%%%%%%%%%%%%%%%%%%%%%%%%%%%%%%%%%%%%%%%%%%%%%%%%%%%%%%%%%%%%
\subsubsection{Corrections to the real part of the gauge coupling function}
\label{sec:corrRealpart}
%%%%%%%%%%%%%%%%%%%%%%%%%%%%%%%%%%%%%%%%%%%%%%%%%%%%%%%%%%%%%%%%%%%%%%%%%%%%%%%

We first calculate the correction to the real part of the gauge coupling function 
from the back-reaction of the $\cG_4$-flux \eqref{eq:G4fluxOnTN} on the warp factor. 
We find this corrected warp factor analytically for the 
full metric \eqref{eq:TN_infty} on the local geometry $\mathcal{Y}_4$ with fluxes $\cG_4$.
Qualitatively, the corrected warp factor then modifies 
all integrals over the internal space $\hat{Y}_4$, in particular \eqref{eq:gaugeKinLeading}, and 
thus corrects the gauge kinetic function. 

The warp factor equation \eqref{Laplaceeq} on $\mathcal{Y}_4$ is given by
\beq \label{wfequation}
\Delta_{\mathcal{Y}_4} e^{3A/2} = *_{\mathcal{Y}_4}(\tfrac12\cG_{4} \wedge \cG_{4}) ,
\eeq
where on the right hand side we have only included the background flux $\cG_{4}$ and 
dropped the remaining terms in \eqref{Laplaceeq}.
In general the precise expression of the the two-forms $\hat{\cF}^I$ on $S_{\rm b}$ 
will induce a non-trivial behaviour of the warp factor on $S_{\rm b}$. 
However, for simplicity we will neglect the non-trivial profile of $\hat{\cF}^I$ on $S_{\rm b}$ 
by averaging over $S_{\rm b}$ as
\beq
\big<\hat \cF^I \wedge \hat \cF^J\big>_{S_{\rm b}} =\delta^{IJ} \frac{1}{\cV_{S_b}}\int_{S_{\rm b}} \hat \cF^I \wedge \hat \cF^I =\delta^{IJ}\frac{n^I}{\mathcal{V}_{S_{\rm b}}},
\eeq
where $\cV_{S_{\rm b}}=\frac12\int_{S_{\rm b}} J\wedge J$ for $J$ denoting the K\"ahler form on 
$S_{\rm b}$. Note that we additionally assumed that the off-diagonal elements $I\neq J$ vanish 
identically. In the brane picture the numbers $n^I$ are then related to the instanton numbers on 
$S_{\rm b}$ in the U($1$) of the $I$-th brane, respectively, as discussed below \eqref{tracessingle}. 
Similarly we average over the 
dependence of the warp factor $e^{3A/2}$ on $S_{\rm b}$ by integrating the right hand side of 
the warp factor equation \eqref{wfequation} over the $S_{\rm b}$. Then we obtain an equation 
between four-forms on $TN^\infty_k$ reading
\beq \label{wfequationSimple}
d *_4d\, e^{3A/2} =\frac{n^I}{2\cV_{S_{\rm b}}} \, \Omega_I^\infty\wedge 
\Omega_I^\infty\, ,
\eeq
where $d$ and $*_4$ denote the exterior derivative respectively the Hodge star on 
$TN_k^\infty$.
In order to solve the warp factor equation \eqref{wfequation} we first evaluate
\beq \label{eq:Omega_IOmega_J}
\Omega^\infty_I \wedge \Omega^\infty_J=\frac{2}{r_{\rm A}^2}V d\Big(\frac{V_I}{V}\Big)\wedge 
(dt+U)\wedge*_{3} d \Big(\frac{V_J}{V}\Big) = -\frac{2}{r_{\rm A}^2}V d\Big(\frac{V_I}{V}\Big)\wedge 
*_{4} d \Big(\frac{V_J}{V}\Big)\,.
\eeq
where we used the relation $*_3dU_I=-dV_I$ and $*_4 dV_I=-(dt+U)\wedge *_3 dV_I$ where the latter follows from
\eqref{eq:vierbeins} and the orientation on $TN_k^\infty$ specified there. Then it is straightforward to show that \eqref{wfequationSimple} is solved by
\beq \label{eq:warpFactor}
e^{3A/2} = 1-\frac{n^I}{2r_{\rm A}^2\cV_{S_{\rm b}}}\Big(\frac{V_I^2}
{V}-V_I\Big)\ ,
\eeq
where we made use of $\Delta_3 V_I\sim\delta(z-z_I)$ on the three-
dimensional base of the Taub-
NUT geometry $TN_k^\infty$ as well as 
\beq \label{eq:V_I/V}
   \frac{V_{I}}{V}({\hat z=\hat z_J,\hat \rho=0}) =  \delta_{IJ}\,.
\eeq

The integration constant in \eqref{eq:warpFactor} is chosen to be $1$ 
to reproduce the 
unwarped case.\footnote{In general the 
precise linear combination of the two solutions to the homogeneous 
equation $d*_4 d g=0$ we have 
to add has to be determined by global boundary conditions on 
$e^{3A/2}$.} With this convention the boundary behavior of the warp factor 
is analyzed as follows. First we introduce a cutoff $M$ ``at infinity'' in the $\hat{\rho}$-direction so that
\beq \label{eq:behaviorInfinity}
	\left.V_I\right\vert_{\hat\rho=M}=0\,,\qquad \left.e^{3A/2}\right\vert_{\hat\rho=M}=1\,.
\eeq
Indeed, this behavior at large $\hat\rho$ is necessary to glue the 
local model $\mathcal{Y}_4$ into a compact Calabi-Yau fourfold 
$\hat{Y}_4$. Then we evaluate the warp factor on the locus 
$\hat{Z}_J:=(\hat{\rho}=0, \hat{z}=\hat{z}_J)$ of one monopole in $TN_k^\infty$. We obtain the warp factor 
\bea \label{eq:behaviorOnBrane}
	\left.(e^{3A/2}-1)\right\vert_{\hat{Z}_J}&=&\Big.\frac{n^I}
	{2r_{\rm A}^2\cV_{S_{\rm b}}}\frac{V_I}{V}\big(1+\sum_{K\neq 
	I}V_K\big)\Big\vert_{\hat{Z}_J}=\Big.\frac{n^I}{2r_{\rm 
	A}^2\cV_{S_{\rm b}}}\big[\delta_{IJ}\big(1+\sum_{K\neq 
	I}V_K\big)+V_I\sum_{K\neq 
	I}\delta_{KJ}\big]\Big\vert_{\hat{Z}_J}\nn \\
	&=& \frac{1}{2r_{\rm A}^2\cV_{S_{\rm b}}}\big[n^J+\sum_{K\neq 
	J}(n^J+n^K)V_K\vert_{\hat{Z}_J}\big]\,,
\eea 
which is finite since the potentials $V_K$ are regular at $\hat{Z}_J$
for $K\neq J$. This result is expected since in the one-monopole case 
the warp factor at the position of the 6-brane should only see the 
localized flux $n^J$ on that brane and fall off to $1$ at distances 
$\hat{\rho}$ far away from the brane. However, we see from 
\eqref{eq:behaviorOnBrane} that in the case of $k$ monopoles, besides 
this back-reaction of the localized flux $n^J$ on the same 6-brane at 
$\hat{Z}_J$ the gravitational back-reaction of the 
localized fluxes $n^K$ from different branes, $K\neq J$, also
affects the warp factor at $\hat{Z}_J$ with a suppression factor 
$V_K\vert_{\hat{Z}_J}$.

Now we are able to calculate the gauge coupling function. This is carried out by considering the kinetic term \eqref{eq:gaugeKinLeading} 
corrected by the warp factor in the general metric ansatz \eqref{warped-background}. Following the same logic as for the flux $\mathcal{G}_4$ in \eqref{eq:G4fluxOnTN} 
we include a three-dimensional field strength in the expansion of $G_4$ as
\begin{equation}
 G_4=\hat{F}^I\wedge \Omega_I^\infty + \cG_4 =F^i\wedge \omega_i^\infty + F^0\wedge\sum_J\Omega^\infty_J+\cG_4 \,,
\label{eq:G4expTN}
\end{equation}
where $I=1,\ldots, k$ and $i=1,\ldots, k-1$.
Then, the three-dimensional gauge fields are embedded into U$(k)$ as 
\begin{equation}
	\hat F^{m} = F^0+F^{m} - F^{m-1}\,,\qquad \hat F^{1} = 	
	F^0+F^{1}\,,\qquad \hat F^{k} = F^0- F^{k-1}\,,
\label{eq:relFields}
\end{equation}
for $m=2,\ldots,k-1$, which is completely analogous to 
\eqref{eq:G4fluxOnTN}.
The three-dimensional  kinetic term for $\hat F^I$ is evaluated in the warped background as 
in section \ref{sec:warpedreduction}, and contains the warped metric \eqref{def-cGw}.
Focusing on the warped metric in the local fourfold $\mathcal{Y}_4$ we obtain
\beq \label{eq:warpedcG}
 \cG_{IJ}^w =  \frac{\cV_w}{4} \int_{\mathcal{Y}_4} e^{3A/2} \Omega^\infty_{I} \wedge *_{\mathcal{Y}_4} \Omega^\infty_{J} 
 = -\frac{\cV_w\cV_{S_{\rm b}}}{4} \int_{TN_k^\infty} e^{3A/2} \Omega^\infty_{I} \wedge  \Omega^\infty_{J} 
\eeq
which is the corrected version of \eqref{cGAB}. Here we used that the Hodge star on $\mathcal{Y}_4$ acts as $\frac{1}
{2}J^2*_4$ and in addition the anti-selfduality \eqref{TNin} of $\Omega_I^\infty$. Noting that the forms 
$\Omega^\infty_I$ are constant over $S_{\rm b}$ we readily integrate out the K\"ahler form to obtain a volume factor 
$\mathcal{V}_{S_{\rm b}}$.
Then we read off the gauge coupling function $\R f_{IJ}$ simply as the coefficient 
of the kinetic term $\hat F^I\wedge * \hat F^J$ in \eqref{eq:C2actionred} from which we see that we have to take 
into account an additional factor of $-2R = -2v^0 / \cV_w$. In addition we note that the Type IIB volume $\R 
T_S=2\mathcal{V}_{\rm S_b}$. Since the warp factor only appears linearly in 
\eqref{eq:warpedcG} we insert the solution \eqref{eq:warpFactor} for  $e^{3A/2}$ to obtain
\begin{equation}
	\Re f_{IJ}\equiv -2\frac{v^0}{\mathcal{V}_w}\cG_{IJ}^w=  \tfrac{1}{4}v^0\R 
T_S\delta_{IJ}-\frac{v^0}{4r_{\rm 
	A}^2}n^K\int_{TN_k^\infty}\Big(\frac{V_K^2}{V}-V_K\Big)\Omega_I^\infty\wedge\Omega_J^\infty\,,
\label{eq:Integralf}
\end{equation}
where we used the property \eqref{TNin} of the $\Omega_I$ on the first term to obtain the proportionality to 
$\delta_{IJ}$.

We immediately recognize the first term in \eqref{eq:Integralf} as 
the leading part of the gauge coupling function \eqref{eq:f_IJCoulomb} on the Coulomb branch 
of the three-dimensional gauge theory. 
The second term in \eqref{eq:Integralf} already resembles the real part 
of the flux induced contribution $\R f_{IJ}^{\text{flux}}$ to the gauge coupling \eqref{eq:fIJD74d} respectively 
\eqref{eq:f_IJCoulomb}. We 
obtain the final expression for the gauge coupling function by 
evaluating the integral in \eqref{eq:Integralf} over the local geometry $TN_k^\infty$. 
However, instead of evaluating this in general,
which is hard due to complicated integrand, we
focus on the weak coupling result $g_s\sim 0$. For small $g_s$, as discussed rigorously in 
appendix \ref{app:TNgeo}, we can use the localization property  
\beq \label{eq:Omega^2weakCoupling}
	\Omega^\infty_I\wedge \Omega_J^\infty\,\rightarrow\,-\frac{1}
	{2\pi}\delta_{IJ}\delta(\hat\rho)\delta(\hat{z}-\hat{z}_I)d\hat t \wedge 
	d \hat \rho \wedge 
	d\varphi \wedge d\hat z\,
\eeq
in local coordinates $\hat{z}$ on the quotient $\mathbb{R}/\mathbb{Z}=S^1$. 
Then we evaluate the integral in \eqref{eq:Integralf} as
\bea
\text{Re}f_{IJ}^{\text{flux}}&= &-\frac{v^0}{4 r_{\rm A}^2}n^K\int_{TN_k^\infty}\Big(\frac{V_K^2}{V}-
V_K\Big)\Omega_I^\infty\wedge\Omega_J^\infty = 
\left.\frac{1}{2}\delta_{IJ}v_0\mathcal{V}_{S_{\rm b}}(e^{3A/2}-1) \right|_{\hat{Z}_I}\nn\\ 
&=& \frac{1}{8}g_s^{-1}\delta_{IJ}\big[n_{IIB}^I+\sum_{K\neq 
	I}(n_{IIB}^I+n_{IIB}^K)V_K\vert_{\hat{Z}_I}\big]\,, \label{eq:Derivedf^flux}
\eea
where we used the evaluation of the warp factor \eqref{eq:behaviorOnBrane} in the last equality and the basic 
relation $\frac{v^0}{r_{\rm A}^2} = g_s^{-1}$ following from \eqref{eq:defcouplings}. Moreover the remaining 
integrals over $\hat{t}$ and $\varphi$ yield a factor $1$ respectively $2\pi$. In addition we identified 
theflux number $n^{K}_{IIB} =2n^K$ due to the orientifolding as noted already in 
\eqref{eq:FtheoryLift}. 

We note, that in the result \eqref{eq:f_IJCoulomb} for $\text{Re}f_{IJ}^{\text{flux}}$
that we obtained by dimensional reduction of the D7-brane effective action to three dimensions 
we only see the first term in \eqref{eq:Derivedf^flux} proportional to $ n^I$. However, this is 
perfectly consistent recalling that $V_K\sim g_s$, cf.~\eqref{correction}, which reveals the 
corrections proportional to $V_K$ in \eqref{eq:Derivedf^flux} as one loop corrections to the 
gauge coupling $f_{IJ}$. These are \textit{not} visible in the string-tree-level D7-brane 
effective action obtained in section \ref{sec:TypeIIBRev}. More precisely the corrections are 
suppressed by $g_s$ and the separation $|\hat{z}_I-\hat{z}_k|$ between the branes as 
\beq \label{eq:V_KatZ_I}
	V_K\vert_{\hat{Z}_I}=\frac{g_s}{4\pi}\Big(\frac{1}{|
\hat{z}_I-\hat{z}_K|} - 2\gamma - \psi( 1 - |
\hat{z}_I-\hat{z}_K|) -\psi(1 + |
\hat{z}_I-\hat{z}_K|)\Big)\,,
\eeq
where we used \eqref{per_Vi} and introduced Euler's constant $\gamma=0.577216\ldots$ as well 
as $\psi(x)$ denoting the digamma function. 
The function $\psi(x)$ is well-defined except at $x\in \{0,-1,-2,\ldots\}$ and since $0<|
\hat{z}_I-\hat{z}_K|<1$, the composition $\psi(1-|\hat{z}_I-\hat{z}_K|)$ is finite
\footnote{In contrast the Poisson re-summed $V_K$ in \eqref{correction} diverges at $\rho=0$, 
though, since Poisson re-summation breaks down for $\rho=0$ and \eqref{correction} is not valid at 
$\rho=0$.}. We note, however, that this implies that the corrections in \eqref{eq:Derivedf^flux} 
diverge as $\frac{1}{|\hat{z}_I-\hat{z}_K|}$ in the case that the branes move on top of each other 
$\hat{z}_K=\hat{z}_I$. Intuitively this is clear since the integral \eqref{wfequationSimple} 
calculates formally the self-energy $E=\int \phi \varrho dV$ of charges in three dimensions by identifying 
$e^{3A/2}$ with the electric potential $\phi$ and $\Omega_I^\infty\wedge \Omega_I^\infty$ with the charge 
density $\varrho$. Thus, by using the approximation \eqref{eq:Omega^2weakCoupling} we formally calculate
the self-energy of a point charge, that is infinite. However,  the self-energy i.e.~the integral 
\eqref{wfequationSimple} is regularized in M-theory by the smooth forms $\Omega_I^\infty$ that smear out the charge 
density $\varrho$.

%%%%%%%%%%%%%%%%%%%%%%%%%%%%%%%%%%%%%%%%%%%%%%%%%%%%%%%%%%%%%%%%%%%%%%%%%%%%%%%
\subsubsection{Corrections to the imaginary part of the gauge coupling function}
\label{sec:corrImaginarypart}
%%%%%%%%%%%%%%%%%%%%%%%%%%%%%%%%%%%%%%%%%%%%%%%%%%%%%%%%%%%%%%%%%%%%%%%%%%%%%%%

In this final section we calculate the flux-induced corrections to the imaginary 
part of the gauge coupling function. These corrections originate from the 11-dimensional
Chern-Simons term $C_3\wedge G_4\wedge G_4$ with an
altered reduction ansatz \eqref{C3_expand} in the presence of a non-trivial
flux $\mathcal{G}_4$. Following the logic of section \ref{sec:warpedreduction} the dependence
of the new three-form $\beta(M^\Sigma)$ on the moduli $M^\Sigma$ of the compactification geometry
is crucial to obtain the coupling $d_{\mathcal{A}\Sigma\Lambda}$ in \eqref{red_warped_action}. 
It is a Chern-Simons term in three dimensions and is identified with the reduction of the 
topological term $\text{Tr}(F\wedge F)$ of the four-dimensional gauge theory to three dimensions in 
\eqref{eq:C2actionred}. We demonstrate this
identification and the reproduction of the right flux correction to the imaginary part of the
gauge coupling and obtain a perfect match in the weak coupling limit 
where we reproduce the flux correction $\sim n^I$ in \eqref{eq:f_IJCoulomb} to the D7-brane gauge coupling.

First we have to identify the appropriate form for the three-form $\beta$ that we defined in \eqref{G4_expand} 
as the Chern-Simons form of the flux $\cG_4=d_8\beta$. From the expansion \eqref{eq:G4fluxOnTN} and recalling 
$\Omega_I^\infty=d_4\eta_I$ we make the ansatz
\beq
	\beta=\cF^I\wedge\eta_I(\varphi_0,\underline{\hat{z}})\,,
\eeq
where we indicated the moduli dependence of $\beta$ on the angle $\varphi_0$ and the position of the $k$ 
periodic monopoles $\underline{\hat{z}}=(\hat{z}_I)$ through the one-forms $\eta_I$. From this it follows
that the relevant terms in the three-dimensional action \eqref{red_warped_action} take the form
\beq \label{eq:warpedActionImPart}
	S_{G_4}^{(3)}\supset 2\pi\int_{\cM_3}\big(d_{IC_0 K}C_0d\hat{z}^K\wedge \hat{F}^I
	+d_{IKC_0}\hat{z}^KdC_0\wedge \hat{F}^I\big)\,,
\eeq
where we identified the RR-axion $\tfrac{k}{2\pi}\varphi_0=C_0$ as before in the definition of the axio-dilaton 
\eqref{eq:identifyC0gs} and set $\tilde{\omega}_I=\Omega_I$ as in \eqref{eq:G4expTN}. Then the coupling 
$d_{IC_0 K}$ is given by
\beq \label{eq:dCouplingOnTN}
	d_{IC_0 K}=-\tfrac{1}{4}\int_{\hat{Y}_4}\Omega^\infty_I\wedge\frac{\partial\beta}{\partial C_0}\wedge
	\frac{\partial\beta}{\partial \hat{z}^K} =-\tfrac{1}
{4}n^J\delta^{JL}\int_{TN_k^\infty}\Omega_I^\infty\wedge\partial_{C_0}\eta_L\wedge\partial_{\hat{z}_K}\eta_J\,.
\eeq
Here we replaced the compact fourfold $\hat{Y}_4$ by our local geometry $\mathcal{Y}_4$ that by its direct 
product structure  $\mathcal{Y}_4=S_{\rm b}\times TN_k^\infty$ allowed us to 
pull out the integral of the flux over $S_{\rm b}$. We note that the two terms in \eqref{eq:warpedActionImPart}
are equal, up to a term proportional to $d(d_{IC_0K})C_0\hat{z}^K\hat{F}^I$, by partial integration and by 
virtue of the antisymmetry of $d_{IC_0K}$ in the last two indices. In general this can yield further subleading 
correction to $\text{Im}f_{IJ}$ that we ignore in the following.

In order to show that \eqref{eq:warpedActionImPart} reproduces the flux correction to the imaginary part of the 
gauge coupling we have to evaluate \eqref{eq:dCouplingOnTN}. This is a lengthy but straight forward calculation. 
Omitting the details we obtain up to exact forms the result
\bea \label{eq:dCouplingEvaluated}
	&\!\Omega_I\!\wedge\!\partial_{C_0}\eta_L\!\wedge\!\partial_{\hat{z}_K}\eta_J\!=\!\frac{1}{2}\left[-\frac{V_J}
	{V}\!
	\left(\frac{V_K}{V}-\delta_{KJ}\right)\!\Omega_L\wedge\Omega_I\!-\!\frac{V_L}{V}\frac{V_K}
	{V}\Omega_J\wedge\Omega_I\!+\!2\frac{V_L}{V}V_K\!\left(\frac{V_J}
	{V}-\delta_{KJ}\right)\!\sum_S\!\Omega_I\wedge\Omega_S\right]&\nn\hspace{-0.5cm}\\
	&\hspace{0Em}+\frac{r_{\rm B}}{r_{\rm A}^2}\frac{V_L}{V}\frac{V_J}
	{V}\left(\frac{V_K}{V}-\delta_{KJ}\right)\left(\Delta V_I-\frac{V_I}{V}\Delta 
	V\right)dt\wedge\hat{\rho}d\hat{\rho}\wedge d\varphi\wedge 
	d\hat{z}\,,&\hspace{-0.5cm}
\eea
where we ommited the superscript $^\infty$ for brevity.
In the derivation we first recall from \eqref{eq:Uwithdvarphi} that $U=r_{\rm A} C_0 d\hat{z}$ and evaluate 
$\partial_{C_0}\eta_L=\frac{V_L}{V}d\hat{z}$ that follows  from \eqref{def-Omegainf}.  
Thus we can drop all terms in $\Omega^\infty_I\wedge \partial_{\hat{z}_K}\eta_J$ which are proportional to 
$d\hat{z}$. 
Next we plug in the definitions for these forms and formally calculate the derivatives in local coordinates. We 
note that due to the dependence of $V_I$ and $U_I$ in \eqref{correction}, \eqref{eq:Aresummed} on only the 
combination $(\hat{z}-\hat{z}_I)$ we can write
\beq
	\partial_{\hat{z}_K}V_I=-\delta_{IK}\partial_{\hat{z}}V_I\,,\qquad 
	\partial_{\hat{z}_K}U_I=-\delta_{IK}\partial_{\hat{z}}U_I\,.
\eeq
Next we write the relation $*_3dU_I=-dV_I$ in local coordinates for the $\varphi$-component $U^\varphi_I$ of 
$U_I$ as
\beq
	r_{\rm B}\hat{\rho}\partial_{\hat{\rho}}V_I=\partial_{\hat{z}}U_I^\varphi\,,\qquad 
	r_{\rm B}\hat{\rho}\partial_{\hat{z}}V_I=-\partial_{\hat{\rho}}U_I^\varphi\,,
\eeq
which is of course in perfect agreement with \eqref{correction}, \eqref{eq:Aresummed},
to recast every term in \eqref{eq:dCouplingEvaluated} as a function of derivatives of $V_I$ and $V$ multiplying 
the top-form $dt\wedge \hat{\rho}d\hat{\rho}\wedge d\varphi\wedge d\hat{z}$. Then we perform partial 
integrations, ignoring boundary terms, until every single partial derivative acts only on fractions $\frac{V_I}
{V}$. Comparing to \eqref{eq:Omega_IOmega_J} in local coordinates, 
\beq
	\Omega_I^\infty\wedge\Omega_J^\infty=-\frac{2r_{\rm B}V}{r_{\rm A}^2}\left[\partial_{\hat{\rho}}\Big(\frac{V_I}
	{V}\Big)\partial_{\hat{\rho}}\Big(\frac{V_J}{V}\Big)+\partial_{\hat{z}}\Big(\frac{V_I}
	{V}\Big)\partial_{\hat{z}}\Big(\frac{V_J}{V}\Big)\right]dt\wedge\hat{\rho}d\hat{\rho}\wedge d\varphi\wedge 
	d\hat{z}\,
\eeq
where we used $*_4d\hat{\rho}=-\hat{\rho}(dt+U)\wedge d\varphi\wedge d\hat{z}$ and $*_4d\hat{z}=-\hat{\rho}
(dt+U)\wedge d\hat{\rho}\wedge d\varphi$ exploiting the vierbein formalism \eqref{eq:vierbeins}, allows us to 
obtain the first two terms in \eqref{eq:dCouplingEvaluated}. However, partial integration in addition produces a 
term
\beq
	\frac{r_{\rm B}}{r_{\rm A}^2}\frac{V_L}{V}V_J\Big(\frac{V_K}{V}-\delta_{KJ}\Big)\Delta_3\Big(\frac{V_I}
	{V}\Big)dt\wedge\hat{\rho}d\hat{\rho}\wedge d\varphi\wedge 
	d\hat{z}\,.
\eeq 
Applying $\Delta_3=*_3d*_3d$ we obtain the last two terms in \eqref{eq:dCouplingEvaluated} and the 
mixed terms with derivatives acting on different terms can be rewritten using
\beq
	\sum_S\Omega^\infty_I\wedge 
	\Omega^\infty_S=\frac{2}{ r_{\rm A}^2V}dV\wedge (dt+U)\wedge*_3 d\Big(\frac{V_I}{V}\Big)=-\frac{2}{ r_{\rm A}^2V}dV\wedge *_4 d\Big(\frac{V_I}{V}\Big)
\eeq
and $*_4dV_I=-(dt+U)\wedge *_3dV_I$ yielding the third term in \eqref{eq:dCouplingEvaluated}.

With the result \eqref{eq:dCouplingEvaluated} we can now evaluate the coupling $d_{IC_0K}$ in 
\eqref{eq:dCouplingOnTN}. According to \eqref{eq:f_IJCoulomb}, \eqref{eq:C2actionred} and 
\eqref{eq:warpedActionImPart} it is related to the imaginary part of the flux correction to the the gauge 
coupling function in the Coulomb branch of $U(k)$ as $\frac{1}{2}\text{Im}f_{IJ}^{\rm flux}$  if we identify
\beq
	\hat{z}_I=\zeta^I=\frac{\xi^I}{r_{\rm B}^2}\,.
\eeq
Again we focus on the extraction of the weak coupling 
behavior $g_s\sim 0$ where the integral \eqref{eq:dCouplingOnTN} for $d_{IC_0K}$ can be evaluated 
explicitly. We recall first the limit \eqref{eq:Omega^2weakCoupling} and note that the potentials $V$, $V_I$ 
obey the Poisson equation \eqref{eq:PoissonV_I}. As in the evaluation of the real part \eqref{eq:Derivedf^flux}
we then replace all four-forms in \eqref{eq:dCouplingEvaluated} by delta-functions and by integration we just 
have to evaluate the different pre-factors at points. Then only the second and 
third term in \eqref{eq:dCouplingEvaluated} contribute yielding $n^J\delta_{JL}\frac{V_L}{V}\frac{V_K}
	{V}\Omega_J\wedge\Omega_I\rightarrow -\delta_{IK}n^I$ respectively
\beq
	n^J\delta_{JL}\frac{V_LV_K}{V}\!\left(\frac{V_J}
	{V}-\delta_{KJ}\right)\!\sum_S\!\Omega_I\wedge\Omega_S\rightarrow\delta_{IK}\big[n^I+n^I\sum_{S\neq 
	I}V_{S}\vert_{\hat{Z}_I}\big]+(\delta_{IK}-1)n^IV_K\vert_{\hat{Z}_I}\,,
\eeq
where it is important for the latter formula to separately consider the cases $I=K$ and $I\neq K$ and to split 
the sum over $J$ into $J\neq I$ and $J=I$.
Thus we obtain the imaginary part of the flux correction to $f_{IJ}$  as
\beq \label{eq:Imffluxcalculated}
	\text{Im}f_{IJ}^{\rm flux} \cong 2d_{IC_0K}=-\tfrac{1}{8}C_0\Big[\delta_{IJ}\big(3 n_{IIB}^I+2\sum_{K\neq 
	I}n_{IIB}^IV_K\vert_{\hat{Z}_I}\big)+2n_{IIB}^IV_J\vert_{\hat{Z}_I}(\delta_{IJ}-1)\Big]\,,
\eeq
where we used as before the definition $\hat{Z}_I:=(\hat{\rho}=0, \hat{z}=\hat{z}_I)$ and the relation 
$n^{K}_{IIB} =2n^K$. Note that the structure is similar to the real part however a precise matching requires to keep 
all terms, most importantly those related to the terms in \eqref{eq:warpedActionImPart} by partial 
integration\footnote{In particular the factor $3$ in \eqref{eq:Imffluxcalculated} arises precisely by partial integration 
and should be cancelled by the omitted terms in \eqref{red_warped_action} that are also obtained by partial 
integration.},  in the reduced action \eqref{red_warped_action}. We emphasize that 
we not only obtain the expected flux correction to the imaginary part of $f_{IJ}$ in \eqref{eq:f_IJCoulomb} 
but also subleading corrections proportional to $g_s$ via $V_J\sim g_s$. These corrections are analogous to the 
to those of the real part in \eqref{eq:Derivedf^flux} and are accordingly identified as one-loop corrections 
that are absent in the strict weak coupling limit and in particular in the the tree-level result 
\eqref{eq:f_IJCoulomb} of the D7-brane gauge coupling. Using the finite expression \eqref{eq:V_KatZ_I} for 
$V_K\vert_{\hat{Z}_I}$ we can predict some of this leading loop correction.

%%%%%%%%%%%%%%%%%%%%%%%%%%%%%%%%%%%%%%%%%%%%%%%%%%%%%%%%%%%%%%%%%%%%%%%%%%%%%%%
%%%%%%%%%%%%%%%%%%%%%%%%%%%%%%%%%%%%%%%%%%%%%%%%%%%%%%%%%%%%%%%%%%%%%%%%%%%%%%%

%%%%%%%%%%%%%%%%%%%%%%%%%%%%%%%%%%%%%%%%%%%%%%%%%%%%%%%%%%%%%%%%%%%%%%%%%%%%%%%
%%%%%%%%%%%%%%%%%%%%%%%%%%%%%%%%%%%%%%%%%%%%%%%%%%%%%%%%%%%%%%%%%%%%%%%%%%%%%%%

\section{Conclusion}
\label{sec:conclusion}

In this work we have studied corrections to the four-dimensional F-theory effective action induced by 7-brane fluxes.
We have argued that this can be done via a three-dimensional M-theory compactification by comparing the result 
to a circle reduction of a genuine four-dimensional $\cN=1$ supergravity theory.  The 7-brane 
fluxes are lifted to M-theory four-form flux $G_4$. The crucial observation was that the $G_4$ flux 
then backreacts and requires a more general Kaluza-Klein Ansatz 
including a non-trivial warp factor and a modified three-form potential. 
The vacuum solutions are the warped Calabi-Yau fourfold backgrounds found in \cite{Becker:1996gj}.  
Thus, the determination of the effective action requires to perform a warped Kaluza-Klein reduction and results 
in new terms depending on the background fluxes. The warping and modified M-theory potential crucially 
depend on the circle direction decompactified in the M-theory to F-theory limit. 
This dependence induces additional terms interpreted as 7-brane flux corrections in four-dimensions.

To explicitly derive these corrections to the effective action it was necessary to solve the warp factor equation and 
give the explicit representatives of the harmonic forms on the internal geometry. Clearly, this requires a knowledge of 
the metric on the internal Calabi-Yau space and is very hard in general. Therefore, we have focused 
on local M-theory geometries, which yield D6-branes at weak coupling. 
More precisely, we considered M-theory 
on Taub-NUT spaces and made use of the explicit 
form of the Taub-NUT metric and its harmonic forms and potentials. For a stack of $k$ D6-branes the 
M-theory background is in fact the multi-centered Taub-NUT space $TN_k$. In order to compare with the 
F-theory action one has to effectively perform a T-duality along a circle to move from Type IIA to Type IIB string 
theory. Hence, it was necessary to introduce an infinite array $TN_k^\infty$ of multi-centered Taub-NUT spaces with 
period given by the circle radius. The metric and harmonic forms have been determined by using a Poisson resummation. 

To determine the warp factor and M-theory three-form potential we have considered the local fourfold 
geometry $S_{\rm b} \times TN_k^\infty$, where $S_{\rm b}$ is a complex surface over which we averaged the solutions.
Remarkably, it was sufficient to use some of the key properties of the $TN_k^\infty$ geometry to solve 
the warp factor equation and give a closed expression of the warp factor in terms of the fluxes and the Taub-NUT potentials
determining its metric. Using this solution in the warped Kaluza-Klein reduction we were able to show that 
the real part of the gauge-coupling function receives a correction quadratic in the flux. In F-theory this corresponds 
to a flux correction to the real part of the gauge coupling function of  a space-time filling 7-brane wrapped on $S_{\rm b}$.
At weak string coupling this additional term precisely yields the flux square correction linear in the inverse Type IIB 
string coupling $\I \tau = 1/g_s$, which can be derived from the Dirac-Born-Infeld action of a D7-brane. To derive the
correction to the imaginary part of the gauge coupling function we had to dimensionally reduce the M-theory 
Chern-Simons term taking into account a back-reacted three-form potential. On our local geometry it was given 
by $\cF^I \wedge \eta_I$, where $\cF^I$ is a two-from flux on $S_{\rm b}$ and $\eta_I$ are the $k$ fundamental 
one-forms on $TN_k^\infty$. While the internal derivative of this correction gives the background flux, its external
derivatives induce the flux-square correction to the 7-brane gauge coupling function. 
Let us note that the classical part of the imaginary part of the 7-brane gauge coupling  arises not from the 
Chern-Simons term of M-theory, but rather from the kinetic term of $G_4$ as a kinetic mixing of U$(1)$ field strengths. 
This is due to the fact, that the imaginary part of the K\"ahler modulus $T_\alpha$ is in fact arising as a 
three-dimensional vector in the M-theory reduction to three dimensions.

In addition to matching the know weak coupling result from the D7-brane action we have shown that 
there extra terms going with a factor $g_s$. It would be desirable to analyze these in more detail. This
would involve solving the integrals over the warp factor corrected gauge coupling without 
employing a sharp localization of $\Omega_I \wedge \Omega_J$. While this is beyond the 
scope of this work, one expects that this can be also done in a closed form using techniques 
known from the study of one-loop integrals.\footnote{See, e.g.~\cite{Angelantonj:2011br,Green:2011vz}, for some 
recent progress in this direction.} A further extension is to include the 
 higher curvature terms in the equation determining the warp factor. These should yield the missing 
 higher curvature terms on the D7-brane action and further complete the four-dimensional effective action 
 of F-theory. This is likely doable in our local geometry where the solution to the warp factor equations 
 can be determined using the explicit metric.
 
 Let us stress that the basic idea presented in this work is much wider applicable. We argued that 
 the M-theory to F-theory lift is subtle, since non-trivial profiles of the fields in the growing 
 extra dimension have to be incorporated with care.  We have shown an example computation for 
 the local geometry near a stack of branes. To perform the evaluation in more general backgrounds 
 appears significantly harder. However, let us point out that in most of our computations only some basic properties 
 of the defining functions were necessary. One expects that one can develop a formalism  
 which does not make use of the metric but rather employs appropriate $\cN=1$ periods of the Calabi-Yau fourfold 
 depending on the complex structure moduli. In the most optimistic scenario, one can use these in a 
 complete warped Kaluza-Klein reduction inducing geometric corrections to all $\cN=1$ characteristic functions 
 determining the four-dimensional effective action.

%%%%%%%%%%%%%%%%%%%%%%%%%%%%%%%%%%%%%%%%%%%%%%%%%%%%%%%%%%%%%%%%%%%%%%%%%%%%%%%%%%%%%
%%%%%%%%%%%%%%%%%%%%%%%%%%%%%%%%%%%%%%%%%%%%%%%%%%%%%%%%%%%%%%%%%%%%%%%%%%%%%%%%%%%%%

\vskip 1cm 
 {\noindent  {\Large \bf Acknowledgements}} 
 \vskip 0.3cm
We would like to thank Yi-Zen Chu, Mirjam Cveti\v c, Monica Guica, Babak Haghighat, Jim Halverson, Albrecht Klemm, 
Hans-Peter Nilles, Eran Palti, Daniel Park, Raffaele Savelli, Stephan Stieberger, Wati Taylor, and Timo Weigand for 
interesting discussions. TG likes to thank the Bethe Center Bonn, UPenn, KITPC, and MIT for hospitality and support 
during the preparation of this work.
DK acknowledges hospitality of the Bethe Center Bonn. DK and MP are also grateful for the hospitality of the MPI Munich.  
The research of TG was supported by a research grant of the 
Max Planck Society. The research of DK was supported by the `Deutsche Telekom Stiftung' and DOE under grant DE-
FG02-95ER40893-A0. The work of MP was supported by the graduate school BCGS, the German National Academic 
Foundation, and the `Deutsche Telekom Stiftung'.

\appendix
%%%%%%%%%%%%%%%%%%%%%%%%%%%%%%%%%%%%%%%%%%%%%%%%%%%%%%%%%%%%%%%%%%%%%%%%%%%%%%%
%%%%%%%%%%%%%%%%%%%%%%%%%%%%%%%%%%%%%%%%%%%%%%%%%%%%%%%%%%%%%%%%%%%%%%%%%%%%%%%

%%%%%%%%%%%%%%%%%%%%%%%%%%%%%%%%%%%%%%%%%%%%%%%%%%%%%%%%%%%%%%%%%%%%%%%%%%%%%%%
\section{Conventions of $\mathcal{N}=1$ actions and dimensionful constants}
\label{app:conventions4dN1}
%%%%%%%%%%%%%%%%%%%%%%%%%%%%%%%%%%%%%%%%%%%%%%%%%%%%%%%%%%%%%%%%%%%%%%%%%%%%%%%

For reference in the main text, let us briefly introduce our conventions for the 
four-dimensional $\mathcal{N}=1$ effective action used in this work. The action 
takes the general form
\beq \label{4dactionconvention}
S^{(4)}_{\cN = 1} = \frac{1}{\kappa_4^2}\int_{\mathbb{R}^{(3,1)}} \!\!\Big(-\frac{1}{2}R * 1 -K_{M\bar{N}}\nabla M^M 
\wedge * \nabla \bar M^{\bar N} -\frac{1}{2}\R f_{AB}F^A\wedge *F^B  - \frac{1}{2}\I f_{AB}F^A\wedge F^B -*V  \Big)\,.
\eeq
Here we introduced the four-dimensional graviational constant, the four-dimensional 
Ricci-scalar $R$, a number of chiral 
superfields with scalar components $M^N$ that are the coordinates of the
K\"ahler manifold of scalar fields with K\"ahler metric $K_{M\bar{N}}
=\frac{\partial^2K}{\partial M^M\partial \bar{M}^{\bar N}}$ and a number of 
vectormultiplets with field strengths $F^A$ with gauge kinetic function
$f_{AB}$  of the chiral multiplets $M^M$. By $*$ we denote the four-dimensional
Hodge star operator and $V$ is the scalar potential that consists of the F-term 
and D-term scalar potential, $V=V_F+V_D$ for
\beq
V_F = e^K\big(K^{M\bar N}D_MWD_{\bar N}\bar W - 3|W|^2 \big), \quad V_D = \frac{1}{2}\R f^{-1 \, AB}D_AD_B .
\eeq
We introduced the superpotential $W$ that is a holomorphic function of the chiral
superfields $M^M$ as well as the $\mathcal{N}=1$ covariant derivative $D_M=\partial_M+K_M$

In the course of deriving this action from String/M-/F-theory it is furthermore useful 
to introduce our conventions for the String, ten- and eleven-dimensional Planck scale as well
as their relation to the D7-brane tension and the four- and three-dimensional Planck scale. These
conventions were originally used in \cite{Grimm:2011tb}
\beq \label{eq:unitsconventions}
 \kappa_{11}^{-2} = \kappa_{10}^{-2} = \kappa_4^{-2} = \kappa_3^{-2} = 2\pi = \mu_7 = T_7\,.
\eeq

%%%%%%%%%%%%%%%%%%%%%%%%%%%%%%%%%%%%%%%%%%%%%%%%%%%%%%%%%%%%%%%%%%%%%%%%%%%%%%%
\section{Linear multiplets and gauge couplings}
\label{app:linMultis+reduction}
%%%%%%%%%%%%%%%%%%%%%%%%%%%%%%%%%%%%%%%%%%%%%%%%%%%%%%%%%%%%%%%%%%%%%%%%%%%%%%%

Let us begin with the dualization of the chiral multiplets with complex scalars $T_\alpha$ into linear multiplets.
More precisely, if $\I T_\alpha$ has a shift symmetry it can be dualized into a two-form $\cC^\alpha_2$, 
which together with $\R T_\alpha$ forms the bosonic components of a linear multiplet \cite{Girardi:1998ju}. To actually 
perform the dualization we collect all terms involving $\I T_\alpha$.
First we turn to the kinetic terms for the $T_\alpha$. These are determined by the four-dimensional K\"ahler potential 
\cite{Grimm:2004uq}
\beq \label{eq:KaehlerpotO7}
	K=-\log(\tau-\bar\tau)-2\log(\mathcal{V}(T + \bar T))\,,
\eeq
where $\mathcal{V}$ is the volume of the Calabi-Yau threefold $Z_3$, which  
considered as a function of $T_\alpha$ is independent of $\tau,\bar  \tau $.
The metric for all complex scalars $M_I=(\tau,T_\alpha)$ is given by 
$K_{I \bar J}$ = $\frac{\partial^2}{\partial M_I \bar \partial \bar{M}_{\bar J}} K$. 
We note that the structure of $K$ at this order implies that there are no kinetic mixing 
terms between $T_{\alpha}$ and $\tau$.

Next we note that in \eqref{action_split} the imaginary part $\I T_\alpha$ also 
appears in front of the theta-angle term $\text{Tr}(F\wedge F)$ in the non-Abelian 
gauge theory. In this case we perform a partial integration and write 
\bea \label{eq:gaugeIm}
S^{(4)}_{\rm gauge, im} &=& - \frac{2\pi}{8}\int_{\cM_4} \delta_S^{\alpha} \, \I T_{\alpha}\  \text{Tr}( F \wedge F) 
     = \frac{2\pi}{8} \int_{\cM_4} \delta_S^{\alpha}\, d \I T_{\alpha} \wedge \omega_{\rm CS}\ ,
\eea
which holds up to a total derivative, and we have defined 
\beq
  \omega_{\rm CS} = A \wedge dA + \tfrac23 A\wedge A \wedge A\ .
\eeq
One can now eliminate $\cG_{\alpha}=d \I T_\alpha$ in favor of its dual $d\cC_2^\alpha$.
We formally achieve this by adding the Lagrange multiplier
\beq
\label{eq:LagrangeMulti}
S^{(4)}_{\rm Lag} =  2\pi\int_{\cM_4}\cG_{\alpha} \wedge d\cC_2^{\alpha} .
\eeq
and eliminate $\cG_{\alpha}$ by its equations of motion. 
First we evaluate the equations of motion yielding
\beq \label{def-GH}
\cG_{\beta} =  - \tfrac{1}{2}  K^{T_{\gamma} \bar T_{\beta}} *  \cH^\alpha_3 \, , \qquad \cH^\alpha_3  =   
d\cC_2^{\alpha} +  \tfrac{1}{8} \delta_S^{\alpha} \omega_{\rm CS}\ , 
\eeq
where we have introduced the modified field strength $ \cH^\alpha_3$.
Then we rewrite the relevant effective action including \eqref{4daction}, \eqref{eq:gaugeIm} and 
\eqref{eq:LagrangeMulti} 
in terms of $\mathcal{G}_\alpha=d \I T_\alpha$ and eliminate $\mathcal{G}_\alpha$ by using \eqref{def-GH}. 
Inserting this into the above action we obtain
\bea \label{eq:C2action}
S_{\cC_2,F}^{(4)} &=& 2\pi\!\!\int_{\cM_4}\!   \tilde K_{\alpha \beta} \, \cH^\alpha_3   \wedge * \cH^\beta_3 
    + \tfrac{1}{4}\tilde K^{\alpha \beta}d\R T_{\alpha} \wedge * d\R T_{\beta} - \tilde K_{\tau \bar \tau}d\tau 
    \wedge * d\bar \tau \nn \\ 
  &&\phantom{2\pi\!\!\int\!} \ \ - \tfrac12 \I f_{\cA \cB}^{\rm flux} F^\cA \wedge F^\cB - \tfrac{1}{2}\R 
  f_{\cA \cB} F^\cA \wedge * F^\cB \ . 
\eea
In order to bring the kinetic term for $\mathcal{C}_2^\alpha$ in the canonical form we have in addition used 
the Legendre-transformed dual K\"ahler potential of \eqref{eq:KaehlerpotO7} given by \cite{Grimm:2004uq}
\begin{equation}
	\tilde{K}(\tau|L)=K+L^\alpha\, \R T_\alpha =\log(\tfrac{1}{6}L^\alpha L^\beta 
	L^\gamma\mathcal{K}_{\alpha\beta\gamma})-\log(\tau-\bar{\tau})\,
\label{eq:LegendreKaehlerpotIIB}
\end{equation} 
for the Legendre-transformed dual variables
\begin{equation}
	L^\alpha=-\frac{\partial K}{\partial \R T_\alpha}=\frac{v^\alpha}{\mathcal{V}}\,,\quad \R T_\alpha= 
	\frac{\partial \tilde{K}}{\partial L^\alpha}
\label{eq:LegendreCoordIIB}
\end{equation}
that was defined in \cite{Grimm:2004uq} to dualize the \textit{real part} $\R T_\alpha$ of the K\"ahler moduli 
to the scalar component of different linear multiplets.\footnote{This is not to be mixed up with the 
dualization of the imaginary part $\I T_\alpha$ performed in this section. In particular, the two-forms 
$D^\alpha_2$ \cite{Grimm:2004uq} forming the linear multiplet together with the $L^\alpha$ are different from 
the two-forms $\mathcal{C}^2_\alpha$ defined in \eqref{eq:LagrangeMulti}.} Essentially we exploited here the 
basic relation 
$K_{T_{\alpha} \bar T_{\beta}} % = -\frac{1}{4}\frac{\partial L^\alpha}{\partial \R T_\beta} 
= -\frac{1}{4}\tilde K^{\alpha \beta}$,
which is an immediate consequence of the general relations of Legendre transformations 
\eqref{eq:LegendreCoordIIB}.

We conclude the discussion of the four-dimensional effective action by noting that $\cC^{\alpha}_2$ has to 
also transform under a non-Abelian gauge transformations $A \rightarrow A+ d\Lambda$ of the vector fields as 
$\cC_2^{\alpha} \rightarrow \cC_2^{\alpha} - \frac{1}{8} \delta_S^{\alpha} \text{Tr}(\Lambda \,  F)$  to 
ensure invariance of $\cH^\alpha_3$ introduced in \eqref{def-GH}. 
Furthermore, the field strength $\cH_2^\alpha$ obeys the Bianchi identity $d \cH_3^\alpha = \frac18 
\delta_S^\alpha\, \text{Tr} (F\wedge F)$.

%%%%%%%%%%%%%%%%%%%%%%%%%%%%%%%%%%%%%%%%%%%%%%%%%%%%%%%%%%%%%%%%%%%%%%%%%%%%%%%
\section{Details of $TN_k$}
\label{app:TNgeo}
%%%%%%%%%%%%%%%%%%%%%%%%%%%%%%%%%%%%%%%%%%%%%%%%%%%%%%%%%%%%%%%%%%%%%%%%%%%%%%%

In this appendix we review some details of the geometry of multi-center 
Taub-NUT space, $TN_k$. We start with the discussion of one monopole, $TN_1$. The metric is given as 
\beq \label{TN}
ds_{TN}^2 = \frac{1}{V}(dt + U)^2 + V d\vec r^2 \, ,
\eeq
where $t$ denotes a periodic coordinate on an $S^1$ and $\vec r$=$(x,y,z)$ three-dimensional Cartesian coordinates 
on $\mathbb{R}^3$. The circle is non-trivially
fibred over $\mathbb{R}^3$.
The function $V$ and the $S^1$-connection $U$ are related by 
\beq \label{eq:conn1}
 *_{3} dU = \pm dV_1\,,
\eeq
where $*_3$ denotes the Hodge star operator on the base $\mathbb{R}^3$ with standard orientation. 
The $\pm$-sign will lead to an self-dual respectively anti-selfdual two-form $\Omega$ as introduced below in 
\eqref{eq:Omegalocal}.
Note that the closedness of $dU$ requires $V=1+V_1$ to be harmonic\footnote{Actually we consider
 fundamental solutions $V$ with $\Delta V_1=\delta^3(\vec{r})$ in the distributional sense.}.
\eqref{eq:conn1} is solved by 
\beq \label{eq:conn1solved}
 V_1 = \frac{r_{\rm A}}{4\pi |\vec{r}|}, \qquad  U  = \pm\frac{r_{\rm A}}{4\pi}\big(-1+ \frac{z}
 {|\vec r|}\big)\frac{x dy - ydx}{x^2 + y^2} = \pm\frac{r_{\rm A}}{4\pi}(-1 + \frac{z}{|\vec{r}|})
  d\varphi ,
\eeq
where we have also introduced cylindrical coordinates\footnote{We note the spherical symmetry of
the one-monopole configuration with $U  = \pm\frac{r_{\rm A}}{4\pi}(-1 + \cos \theta) d\varphi$ in
spherical coordinates. We use cylinder coordinates to prepare for the discussion of appendix
\ref{app:TNchaingeo}.} with $|\vec{r}|=\sqrt{\rho^2+z^2}$ for $\rho\in \mathbb{R}_+$,
$\varphi\in[0,2\pi]$, $z\in\mathbb{R}_+$. $r_A$ can be thought of as the charge of the monopole,
but more importantly in our context is its interpretation as the circumference of the  
$S^1$-fibre at infinity, as discussed below. We note that the term $\mp d\varphi$ in $U$ is an
integration constant, that is not fixed by the condition $*_3dU=\pm dV_1$ but by the condition of
smoothness of $U$, i.e.~the absence of a Dirac string.
Indeed, the one-form $U$ in \eqref{eq:conn1solved} is only a local one-form representing the 
global connection of the 
Dirac monopole in a coordinate patch. To see this note
the presence of the Dirac string, that is the locus where the local expression $U$ is not 
well-defined. In cylindrical coordinates since $d\varphi$ is not 
well-defined at $\rho=0$, in order to have a well-defined one-form containing $d\varphi$ 
its pre-factor has to vanish on the locus $\rho=0$. However the pre-factor of
$d\varphi$ in the one-form $U$ in \eqref{eq:conn1solved} vanishes only on the positive
$z$-axis for the choice of integration constant $-1$ and $U$ is
ill-defined on the negative $z$-axis. This is precisely the Dirac string. 
Thus, $U$ is only a local one-form well-defined only on the positive $z$-axis and one 
has to introduce at least one further patch and with another local one-form that is required to 
be well-defined on the negative $z$-axis. Thus, one introduces the two patches $\mathcal{U}_\pm$ 
and the corresponding connections denoted $U^\pm$ reading 
\bea \label{eq:2Patches}
 \mathcal{U}_+&=&\{(r,\varphi,z)\,\vert\,0\leq z\}\,:  \quad U^+ =\pm \frac{r_{\rm A}}{4\pi}(-1+\frac{z}
 {\sqrt{\rho^2+z^2}})d\varphi \,,  \nn \\
 \mathcal{U}_-&=&\{(r,\varphi,z)\,\vert\,z\leq 0\}\,: \quad U^- = \pm\frac{r_{\rm A}}{4\pi}(1+\frac{z}
 {\sqrt{\rho^2+z^2}})d\varphi , 
\eea
that differ only by the integration constant in the $d\varphi$-component. They are related by the gauge
transformation $U^+=U^-\mp\frac{r_A}{2\pi}d\varphi$. In particular we note that $U^\pm$ vanishes 
precisely on the positive (negative) $z$-axis and thus can be glued together to form a smooth 
global gauge connection.

Consequently, in order for the metric to be gauge invariant, i.e.~the term $dt+U$ to be globally
defined, the coordinate $t$ has to compensate this gauge transformation and cannot be globally
defined either. Thus we have to introduce two coordinates $t^{\pm}$ on $\mathcal{U}_{\pm}$ that
are related by the gauge transformation 
\beq \label{eq:periodicityT}
t^+ = t^- \pm \frac{r_{\rm A}}{2\pi}\varphi \quad \Rightarrow \quad t \sim t + r_{\rm A},
\eeq
where the sign $\pm$ again refers to the choice in \eqref{eq:conn1} and we inferred the periodicity of $t$ as $\varphi$ 
is identified modulo $2\pi$ \cite{Misner:1963fr}. Thus we see
that the parameter $r_{\rm A}$ sets the circumference of the $S^1$ at infinity, 
$|\vec{r}|\rightarrow \infty$, as the potential $V\rightarrow 1$ in the metric \eqref{TN}. It is 
important to emphasize that only with this circumference we have a globally well-defined $S^1$-
fibre radius.

Next we comment on the smoothness of $TN_1$, where we assume $*_3 dU=+dV_1$ for this paragraph
to avoid confusion. In fact the singularity of $V$ at the origin is 
just a coordinate singularity.
In spherical coordinates one  can expand the metric
\eqref{TN} around the origin using $V\sim V_1$ and the coordinate transformation $q^2=|\vec r|$,
$t^\pm=-\frac{r_A}{4\pi}(\psi\pm \varphi)$ to identify it near the origin as the flat metric 
on $\mathbb{R}^4$ iff $\psi$ has period $4\pi$. This metric is obviously smooth. We note that in 
the case of multiple monopoles $TN_k$ discussed next, the space is still smooth for generic 
positions of the $k$ monopoles, however, develops a deficit angle $2\pi/k$, i.e.~locally becomes
$\mathbb{R}^4/\mathbb{Z}_k$, for $k$ coincident monopoles.

We conclude the analysis of the $TN_1$ geometry by analyzing its (co)homology.
Depending on the sign in \eqref{eq:conn1} $TN_1$ admits a selfdual (sign $+1$) respectively anti-selfdual 
(sign $-1$) two-form that is locally given by 
\beq \label{eq:Omegalocal}
\Omega = d\eta = \frac{1}{r_{\rm A}}d\Big(\frac{V_1}{V}\big(dt + U \big) - 
U \Big) %= \frac{1}{4\pi(r+\frac{r_{\rm A}}{4\pi})^2}\big(e^0 \wedge e^r - e^{\theta} 
%\wedge e^{\varphi}\big) 
\, .
\eeq
As the one-form $U$ is not globally defined as pointed out in \eqref{eq:2Patches}, the 
one-form $\eta$ in turn is not a global form and thus $\Omega$ is not a globally exact form. On
the two patches $\mathcal{U}_{\pm}$ the two local one-forms denoted $\eta^\pm$ are given by
inserting $U^\pm$ defined in \eqref{eq:2Patches} into \eqref{eq:Omegalocal} yielding 
\beq \label{eq:eta+-}
	\eta^\pm=\frac{1}{r_A}\left(\frac{V_1}{V}(dt+U)-U^\pm\right)\,,
\eeq
where we used that the term $dt+U$ is a global one-form by virtue of \eqref{eq:periodicityT}.
It further holds the normalization\footnote{The sign of the $\Omega^2$ can be obtained
for any (anti-)selfdual $\Omega$ since $\Omega \wedge \Omega$ = $\pm\Omega \wedge*
\Omega$, but $\int \Omega \wedge* \Omega$ positive (negative). We can also switch between a self-dual and anti-
selfdual form by changing the orientation on $\mathbb{R}^3$.}
\beq
\int \Omega \wedge \Omega = \pm 1.
\eeq
Furthermore we note the limit\footnote{To prove that we use the following mathematical
statement. Let $(f_j)_{j \in \mathbb{N}}$ be a sequence of positive functions 
defined on $\mathbb{R}^n$, s.t. $\int_{\mathbb{R}^n} f_j(x) dx$ = 1 
$\forall$ $j$. Furthermore $f_j$ converges uniformly to zero on any 
set 0 $<$ $a$ $<$ $|x|$ $<$ $1/a$, for any $a$ $>$ 0, then $f_j$ 
$\rightarrow$ $\delta$ in the distributional sense. This can be 
seen by recalling that uniform convergence means convergence in the 
maximum norm and it is 
easy to see that 
\beq
\underset{|\vec{r}| \in [a, \infty]}{\text{max}} \frac{\frac{r_{\rm A}}{2\pi} r}{(|\vec{r}|+ \frac{r_{\rm A}}{4\pi})^3} 
= \frac{\frac{r_{\rm A}}{2\pi} a}{(a+ \frac{r_{\rm A}}{4\pi})^3} \overset{r_{\rm A}
 \rightarrow 0}{\longrightarrow} 0,
\eeq
which establishes the desired result. }  
\beq \label{delta}
\Omega \wedge \Omega \rightarrow \pm\frac{1}{2\pi} \delta(\rho)\delta(z)d\tilde t \wedge d\rho \wedge 
 d\varphi \wedge dz , \quad \text{for} \quad r_{\rm A} \rightarrow 0\,,
\eeq
where we have introduced a new coordinate $\tilde t$ by $\tilde t$ $=$ $t/r_{\rm A}$. 
That identifies $\Omega\wedge\Omega$ as the dual of the origin in 
$\mathbb{R}^3$ for $r_{\rm A} \rightarrow 0$.

The results of the one-monopole geometry carry easily over to the multi-center case,
denoted $TN_k$. For this one makes the multi-center
ansatz
\beq 
V = 1 + \sum_{I=1}^k V_I, \quad U = \sum_{I=1}^k U_I,\quad V_I 
= \frac{r_A}{4\pi{|\vec r-\vec{r}_I|}}, \quad *_{3} d U_I=dV_I \,,
\eeq
where $\vec r_I$ denote the positions of the $k$ monopoles. The connection $U$ is defined as the
sum of gauge connections $U_I$ constructed for each monopole $I$ along the lines of 
\eqref{eq:2Patches}. To write down an expression for the connection $U$ in local coordinates is 
a bit subtle due to the dependence of the integration constant in $dV_I=*_{\mathbb{R}^3}dU_I$ on 
the choice of coordinate patches covering $TN_k$. As in \eqref{eq:2Patches} we have use two 
patches around each of the $k$ monopoles with corresponding local one-forms $U_I^\pm$ in order 
to avoid a corresponding Dirac string. Placing the $I$-th monopole at the origin, we identify
$U_I^\pm=U^\pm$ as defined in \eqref{eq:2Patches}. Then, in writing down $U=\sum_I U_I$ at a
given point on $TN_k$ we have to decide for each connection $U_I$ separately to either use 
the local one-form $U_I^+$ with integration constant $-1 \cdot d\varphi$ or $U_I^-$ with $1 
\cdot d\varphi$. Thus, adding up the respective integration constants of the $U_I$ the 
integration constant in the local expression for $U$ can take any 
value between $-k\cdot d\varphi$ and $k\cdot d\varphi$ depending on the point on $TN_k$.
\footnote{To illustrate this further, let
us define a patching of $TN_k$ by drawing $k$ two-dimensional planes in $\mathbb{R}^3$ through
each of the $k$ monopoles so that no other monopole is contained in the same plane. For each
monopole this defines a partial order by what we call ``above'' and ``below'' the corresponding
plane in $\mathbb{R}^3$ and we accordingly assign $U_I^\pm\cong U^\pm$. Then for every point in
$TN_k$ we know whether it lies above or below the $I$-th plane and can thus write down the local
expression for $U$ by adding up the integration constants $\mp 1$ of the individual $U_I^\pm$.}

In contrast, the combination $dt+U$ is again unique since it is globally well
defined by virtue of the condition \eqref{eq:periodicityT} around each individual monopole. 
This then implies that in order to get a smooth solution all monopoles have to have the same
charge $r_{\rm A}$. 

The multi-center solution $TN_k$ admits $k$ two-forms locally defined by
\beq \label{twoforms}
\Omega_I = d\eta_I = \frac{1}{r_{\rm A}}d\Big(\frac{V_I}{V}(dt 
+ U) - U_I\Big)\,,
\eeq 
where the two different signs in $*_3dU_I=\pm dV_I$ yield (anti-)selfduality.
They obey the relation
\beq \label{eq:deltaij}
\int_{\mathbb{R}^3 \times S^1} \Omega_I \wedge \Omega_J = \pm \delta_{IJ} .
\eeq
Indeed, we can choose coordinates such that the $I$-th monopole is centred at the origin and
that the two-plane $z=0$ does not contain a different, $K$-th monopole, $K\neq I$.\footnote{We
demand that the plane $z=0$ contains no other monopole although both $\Omega_I$ and $\eta_I$ are 
well-defined at $\vec{r}=\vec{r}_K$.}
This allows us to identify $V_I$ and $U_I$ with the one-monopole connection of $TN_1$ in 
\eqref{eq:conn1solved}. 
Then we introduce spherical coordinates and the coordinate patches of \eqref{eq:2Patches}
and identify $U_I^\pm\equiv U^\pm$.
Since the coordinate patches $\mathcal{U}_\pm$ are just the upper and lower halfspaces of
$\mathbb{R}^3$, $z\leq0$ respectively $z\geq0$, they share, though with opposite
orientation, the common boundary $H$ given by
\beq
	H = \{(r, \varphi,z=0)\}\,.
\eeq
By virtue of Stokes' theorem we may pull the integral of any exact form to this boundary
$H$.
Then we evaluate \eqref{eq:deltaij} taking into account the opposite orientation of $H$,
\beq \label{eq:omegaIJCalcTN1}
\int \Omega_I \wedge \Omega_J  = \!\!\int_{S^1_t \times H} (\eta^+_I - \eta^-_I) \wedge
\Omega_J= \pm\int_{S^1_t \times H} \frac{1}{2\pi} d\varphi \wedge \Omega_J= \pm\int_{S^1_t}\int_{0}^\infty \frac{1}
{r_{\rm A}}d \frac{V_J}{V} \wedge dt = \pm\left.
\frac{V_J}{V} \right|_{\rho=0}^{\infty} \!\!\!= \pm \delta_{IJ}\,,
\eeq
where we first used \eqref{eq:eta+-} with \eqref{eq:2Patches}  and then integrated $dt$ over 
$S^1_t$. In the
last step we exploited that $V_J/V$ vanishes at $\rho=\infty$, as $V\rightarrow 1$ while
$V_1\rightarrow 0$, and vanishes at $\rho=0$ as well except when $V_J=V_I$ yielding
$V_I/V=1$, since the pole $V_I\rightarrow \infty$ cancels precisely the pole $V\rightarrow
\infty$. 

We note that the area of the two-cycles $S_i$ spanning $H_2(TN_k,\mathbb{Z})$ introduced
in \eqref{eq:S_idef}  reads
\beq \label{volume}
\int_{S_i} \text{vol}_{S_i} = \int_{S^1} \int_{\vec{r}_i}^{\vec{r}_{i+1}} V^{\frac{1}{2}}
V^{-\frac{1}{2}} 
= r_{\rm A}|\vec{r}_{i} - \vec{r}_{i+1}|.
\eeq
The forms $\hat{\omega}_i$ = $\Omega_i$ - $\Omega_{i+1}$, $i=1,\ldots, k-1$, spanning its
Poincare dual %$H^2_{\text{cpct}}(TN_N,\mathbb{Z})$
fulfill the following conditions
\beq 
\int \hat{\omega}_i \wedge \hat{\omega}_j = \pm C_{ij}, \quad \int_{S_i} \omega_j =\pm C_{ij} \,,
\eeq
again depending on (anti-)selfduality of $\Omega_I$.
The first statement is clear due to \eqref{eq:deltaij}. For the second 
one we calculate
\beq
\int_{S_i} \omega_j = \int_{\partial S_i} \eta_j - \eta_{j+1} 
= \pm\Bigg(\left. \frac{V_j}{V}\right|_{\vec{r}_{i}}^{\vec{r}_{i+1}} 
- \left. \frac{V_{j+1}}{V}\right|_{\vec{r}_{i}}^{\vec{r}_{i+1}}  \Bigg).
\eeq

%%%%%%%%%%%%%%%%%%%%%%%%%%%%%%%%%%%%%%%%%%%%%%%%%%%%%%%%%%%%%%%%%%%%%%%%%%%%%%%
\section{Details of $TN_k^\infty$}
\label{app:TNchaingeo}
%%%%%%%%%%%%%%%%%%%%%%%%%%%%%%%%%%%%%%%%%%%%%%%%%%%%%%%%%%%%%%%%%%%%%%%%%%%%%%%

The metric of infinitely many Kaluza-Klein monopoles placed with equal spacing $r_{\rm B}$
along a straight line in $\mathbb{R}^3$ is again of the from \eqref{TN}.
Moreover due to the cylinder symmetry of the set-up it is convenient to introduce cylindrical 
coordinates $\rho$ = $\sqrt{x^2 +y^2}$, $\varphi$ = $\arctan(y/x)$ and $z$ being a coordinate on 
the axis along which the monopoles are aligned.  After forming the quotient $z\sim z+r_{\rm B}$ 
we denote this space by  $TN_1^\infty$. The potential $V$ reads
\beq  \label{eq:formsTN1}
V = 1+ \frac{r_{\rm A}}{4\pi}\left(\sum_{\ell \in \mathbb{Z}}\frac{1}{\sqrt{\rho^2 + 
(z+ \ell r_{\rm B})^2}} - \sum_{\ell \in \mathbb{Z}^*} \frac{1}{r_{\rm B}|\ell|}\right)\,.
\eeq
We note that $V$ is now a harmonic function\footnote{Again we have $\Delta_3
V=\delta^3(\vec{r})$ in the distributional sense.} on $\mathbb{R}^2$ $\times$ $S^1$ due to
the periodicity along the $z$-axis. Thus we can view the geometry of $TN_1^\infty$ as a
single Kaluza-Klein monopole on $\mathbb{R}^2\times S^1$, treated as an image charge
problem on $\mathbb{R}^3$. 
The last term in \eqref{eq:formsTN1} is a regulator that assures the convergence of the sum. 
Note that the
precise form of the regulator can be modified by any finite constant.
The corresponding connection $U=\sum_I U_I$ with $*_3dU_I=dV_I$ is given on the patch $z\in[0,r_B[$ as 
\beq \label{eq:UTN^Infinity}
U = \frac{r_{\rm A}}{4\pi}\left(-1+\sum_{\ell \in \mathbb{Z}} \frac{z - \ell r_{\rm B}}{\sqrt{\rho^2 + (z- \ell r_{\rm 
B})^2}}\right)
d\varphi\,,
\eeq
where $-1\cdot d\varphi$ is a choice of integration constant so that $U$ is regular on 
$[0,r_B[$. In fact, treating $TN_1^\infty$ as an image charge problem there is a Dirac string 
for every monopole at $\vec{r}_I=(0,0,\ell r_B)$ as in appendix \ref{app:TNgeo}. Again 
$d\varphi$ is ill-defined for $\rho=0$ and so is $U$ unless the coefficient of $d\varphi$ 
vanishes. Evaluating $U$ in \eqref{eq:UTN^Infinity} at $\rho=0$ we have chosen our 
regularization such that for $z\in [0,r_B[$,
\beq
\left.\sum_\ell\frac{z - \ell r_{\rm B}}{\sqrt{\rho^2 + (z- \ell r_{\rm B})^2}}\right\vert_{\rho=0}= \sum_\ell\text{sign}
(z-\ell r_B)=1
\eeq
and $U=0$,  i.e.~well-defined. However, when considering for instance $z\in[r_B,2r_B[$ we 
evaluate, in the 
same regularization $\sum_\ell \text{sign}(z-\ell r_B)=3$ and the one-form $U$ in 
\eqref{eq:UTN^Infinity} is ill-defined.
Thus, we introduce patches $\mathcal{U}_n$, $n$ integer, that cover the $z$-axis in increments 
of $r_B$ and local one-forms $U^n$,
\beq \label{eq:U^n}
\mathcal{U}_n=\{(\rho,\varphi,z)\,|\,n r_B\leq z<(n+1)r_B\}\,:\qquad U^n=\frac{r_{\rm A}}{4\pi}\left(-1-2n+\sum_{\ell 
\in \mathbb{Z}} \frac{z - \ell r_{\rm B}}{\sqrt{\rho^2 + (z- \ell r_{\rm B})^2}}\right)
d\varphi\,.
\eeq
The $U^n$ are well-defined on $\mathcal{U}_n$ and related by the gauge transformation 
$U^{n+1}=U^n-\frac{r_A}{2\pi}d\varphi$.
In other words, when crossing the lines $z=nr_B$ from below (above) we have to change the
integration constant in the local one-form by $-2d\varphi$ ($+2d\varphi$). It is important to 
note, that $U$ in \eqref{eq:UTN^Infinity} descends to a one-form which is well-defined along 
the whole $S^1$ of the compactified $z$-direction, $z\sim z+r_B$. 

In order to get a better understanding of $V$ and $U$ we perform a Poisson resummation
of these two quantities. Recall that a Poisson resummation relates a function $f$ of
period one and its Fourier-transform $\hat f(k)=\int_{-\infty}^{\infty}f(x)e^{-2\pi i
kx}dx$ via \cite{Grafakos}
\beq \label{Poisson}
\sum_{k \in \mathbb{Z}} \hat f (k) e^{{2\pi i k x}} = \sum_{k \in \mathbb{Z}} f(x + k) .
\eeq
The Fourier-transform of $f(z)$ $=$ $\frac{1}{\sqrt{\rho^2 + z^2}}$ is $\hat f(k)$ =
$2K_0(2\pi\rho k)$, which is the zeroth modified Bessel function of second kind and
shows the following asymptotic behaviour near zero,
\beq \label{Kasym}
K_0(x) = - \log \frac{x}{2} - \gamma, \quad x  \rightarrow 0, \qquad \gamma =
\underset{N \rightarrow \infty} \lim \sum_{k=1}^{N} \frac{1}{k} - \log N \, ,
\eeq
where $\gamma$ is the Euler-Mascheroni constant.
We now plug $f(z)$ $=$ $\frac{r_{\rm A}}{4\pi r_{\rm B}}\frac{1}{\sqrt{\hat \rho^2 +
\hat z^2}}$ with $\hat \rho=\frac{\rho}{r_{\rm B}}$ and $\hat z=\frac{ z}{r_{\rm B}}$
as well as $\hat f(k)$ = $\frac{r_{\rm A}}{2\pi}K_0(2\pi\hat \rho |k|)$ into \eqref{Poisson} and obtain
\beq
V = 1+\frac{r_{\rm A}}{4\pi r_{\rm B}}\Big(\sum_{\ell \in \mathbb{Z}} \frac{1}{\sqrt{\hat \rho^2 + (\hat z + \ell)^2}} - 
\sum_{\ell \in \mathbb{Z}^*} \frac{1}{|\ell|}\Big) = 1+\frac{r_{\rm A}}{2\pi r_{\rm B}}\Big(\sum_{\ell \in \mathbb{Z}^*} 
K_0(2\pi \hat \rho |\ell|) e^{2\pi i \ell \hat z} + K_0(0) - \sum_{\ell>0} \frac{1}{\ell}\Big).
\eeq
The right hand side contains two divergent terms, $K_0(0)$ and $\sum_{\ell >0} \frac{1}
{\ell}$. We therefore have to take a suitable limit to get a finite result by considering
and calculating, using \eqref{Kasym}, 
\beq
\frac{r_{\rm A}}{2\pi r_{\rm B}}\underset{N \rightarrow \infty} \lim \Big( K_0 \big(\frac{2\pi\hat \rho}{N}\big) - 
\sum_{\ell =1}^{N} \frac{1}{\ell}\Big) = \frac{r_{\rm A}}{2\pi r_{\rm B}}\underset{N \rightarrow \infty} \lim\Big(- \log 
\frac{\pi \hat \rho}{N} -\gamma -  \sum_{\ell=1}^{N} \frac{1}{\ell}\Big) = -\frac{r_{\rm A}}{2\pi r_{\rm B}}\log(\frac{\hat 
\rho}{\Lambda}),
\eeq
where $\Lambda$ comprises all constants including an eventually shift in the regulator term. For the concrete regulator 
in \eqref{PoissonV} we have $\Lambda=1/(\pi e^{2\gamma})$. Finally we obtain 
\beq \label{finalV}
V = 1+\frac{r_{\rm A}}{4\pi r_{\rm B}}\Big(\sum_{\ell \in \mathbb{Z}} \frac{1}{\sqrt{\hat \rho^2 + (\hat z + \ell)^2}} - 
\sum_{\ell \in \mathbb{Z}^*} \frac{1}{|\ell|}\Big) = 1 - \frac{r_{\rm A}}{2\pi r_{\rm B}}\Big(\log\frac{\hat \rho}{\Lambda} 
- 2\sum_{\ell>0} K_0(2\pi\hat \rho \ell) \text{cos}(2\pi  \ell \hat z)\Big) .
\eeq
Similarly one can also perform a Poisson resummation for the connection $U$, which is given by
\beq
U = \frac{r_{\rm A}}{4\pi}\Big(-1+\sum_{\ell \in \mathbb{Z}} \frac{(\hat z - \ell)}{\sqrt{\hat \rho^2 + (\hat z - 
\ell)^2}}\Big)d\varphi\,,
\eeq
for $0\leq \hat{z}<1$.
Using that the Fourier transform of $f(\hat{z})= \frac{\hat{z}}{\sqrt{\hat{\rho}^2 + \hat{z}^2}}$ reads $\hat f 
(k)=2i\hat{\rho} \text{sign}(k) K_1(2\pi \hat{\rho} |k|)$ we can perform a Poisson resummation for the connection 
as well, finding naively
\beq
U = \frac{r_{\rm A}}{4\pi}\Big(-1+ 2i\hat \rho \sum_{\ell \in \mathbb{Z}} \text{sign}(\ell) K_1(2\pi \hat \rho |
\ell|)e^{2\pi 
i \ell \hat z}\Big)d\varphi \, .
\eeq
Note that the contribution $\ell$ = 0 is again ill defined. We recall that 
\beq
K_1(x) \sim \frac{1}{x}, \qquad x \ll 1.
\eeq
This enables us to regularize the $\ell=0$ contribution, i.e.~$\text{sign}(\ell)$, as
\beq
\underset{\ell \rightarrow 0}{\text{lim}}\, \frac{1}{2}\Big(i\hat \rho \frac{1}{2\pi \hat \rho \ell}(1+2\pi i \ell \hat z) - 
i\hat \rho \frac{1}{2\pi \hat \rho \ell}(1-2\pi i \ell \hat z) \Big) = - \hat z\,.
\eeq 
We finally obtain
\beq
U = - \frac{r_{\rm A}}{4\pi}\Big(1+2\hat z + 4\hat \rho \sum_{\ell >0}  K_1(2\pi \hat \rho\ell)\text{sin}(2\pi  \ell \hat 
z)\Big)d\varphi \, .
\eeq
Note that this is cohomologically equivalent by adding sterm proportional to $d(\hat z\varphi)$ and $d\varphi$ yielding 
\beq
U = \frac{r_{\rm A}}{2\pi} \varphi d \hat z + \frac{r_{\rm A}}{2\pi} \Big(-2\hat \rho 
\sum_{\ell >0} K_1(2\pi \hat \rho \ell)\text{sin}(2\pi  \ell \hat z)\Big)d\varphi \, .
\eeq

As in the non-periodic case, one can easily generalize to the multi-center case $TN_k^\infty$. 
We restrict ourselves to the case that all monopoles are located at $\big(\hat{\rho}=0, \hat z = \hat 
z_I\big)_{I= 1, ...,  k}$, i.e. we consider $k$ periodic chains of monopoles that are shifted 
among each other. The corresponding re-summed potentials and connections are given for $I=1,
\ldots, k$ and $\hat{z}\in [\hat{z}_I,\hat{z}_I+1[$ by
\bea
V_I &=&  - \frac{r_{\rm A}}{2\pi r_{\rm B}}\Big(\log\frac{\hat \rho}{\Lambda} -2 \sum_{\ell>0} 
K_0\big( 2\pi \hat \rho \ell \big) \text{cos}(2\pi  \ell (\hat z -\hat z_I))\Big)\,,\\
U_I &=& - \frac{r_{\rm A}}{4\pi}\Big(1+2(\hat z-\hat{z}_I) + 4\hat \rho \sum_{\ell >0} K_1(2\pi 
\hat \rho \ell)\text{sin}(2\pi  \ell (\hat z-\hat{z}_I)\Big)d\varphi \, ,
\eea
that obey $*_3dU_I=-dV_I$.
Generalizing the patches of \eqref{eq:U^n} to $k$ monopoles as 
\beq \label{eq:patchU_n(I)}
\mathcal{U}_n(I)=\{(\hat{\rho},
\varphi,\hat{z})\,|\,n +\hat{z}_I\leq \hat{z}<\hat{z}_I+n+1\}\,,
\eeq
we can construct local one-forms $U_I^n$ for other values of $\hat{z}$ by changing the integration constant by $\pm 
2$. In direct analogy with \eqref{eq:U^n} they read on $\mathcal{U}_n(I)$ as
\beq \label{eq:U_I^n}
	U_I^{n}=-\frac{r_A}{4\pi}(1+2n+2(\hat z-\hat{z}_I) + 4\hat \rho \sum_{\ell >0} K_1(2\pi 
	\hat \rho \ell)\text{sin}(2\pi  \ell (\hat z-\hat{z}_I))\Big)d\varphi\,,
\eeq 

Analogously to \eqref{twoforms} the space $TN_k^\infty$ also exhibits $k$ anti-self-dual two-forms
given by
\beq \label{twoformsinfty}
\Omega^\infty_I = d\eta_I = \frac{1}{r_{\rm A}}d\Big(\frac{V_I}{V}(dt 
+ U) - U_I\Big)\, .
\eeq 
The expression for the local one-forms $\eta_I$ depends on the coordinate patches 
$\mathcal{U}_n(I)$, i.e.~the value of $\hat{z}$, through the dependence of the 
$U_I^n$ in \eqref{eq:U_I^n} on the coordinate patch. The local one-forms are denoted $\eta_I^n$. 
The combination $(dt+U)$ for $U=\sum_I U_I$ is again globally defined by appropriately 
defining local coordinates $t$. 
 
We would like to check that the relation
\beq
\int \Omega^\infty_I \wedge \Omega^\infty_J = - \delta_{IJ}
\eeq
still holds in the periodic case. First we center the $I$-th monopole at the origin 
$(\hat{\rho},\varphi,\hat{z})=0$. Then we use as in the one monopole case 
\eqref{eq:omegaIJCalcTN1} the exactness of $\Omega^\infty_I$ on the patches $\mathcal{U}_n(I)$ 
of \eqref{eq:patchU_n(I)}. Since we eventually work on the quotient $\hat{z}\sim \hat{z}+1$ we 
integrate over the interval $\hat{z}\in[0,1]$, but have to keep in mind that the integration 
constant in $U_I$ jumps by $-2d\varphi$ when $\hat{z}\rightarrow 1$ from below.
As mentioned earlier the boundaries of $\hat{z}\in[0,1]$ representing $S^1$ are simply
\beq
H = \{(\hat \rho, \varphi,z=0)\}\,,
\eeq
with opposite orientation, respectively. We readily perform the pullback of the integral by 
Stokes theorem as 
\beq \label{interstep}
\int \Omega_I \wedge \Omega_J=  \int_{S^1_t \times H}  (\eta^1_I-\eta^0_I)\wedge\Omega_J 
=\int_{S^1_t\times H}\frac{1}{2\pi}d\varphi\wedge\Omega_J=\int_{S^1_t}\int_{0}^\infty 
\frac{1}{r_{\rm A}}d \frac{V_J}{V} \wedge dt =\left.\frac{V_J}
{V}\right\vert_{\hat{\rho}=0}^\infty=-\delta_{IJ}\, .
\eeq
Here we used the local expression \eqref{eq:U_I^n} and \eqref{twoformsinfty} to evaluate 
$\eta_I^1-\eta_I^0\sim 2d\varphi$ in the second equality and exploited the behaviour of $V_J/V$ at 
$\hat{\rho}=0,\infty$ as for $TN_1$ to obtain the last equality.

We conclude by representing any metric of the form \eqref{TN} in terms of Vierbeins $e_i$ \cite{Misner:1963fr}
\beq \label{eq:vierbeins}
e^0 = \frac{1}{\sqrt{V}}\big(dt + U \big), \quad e^i = \sqrt{V}dx^i, \quad i = 1,2,3 .
\eeq
Vierbeins make it particularly easy to evaluate the Hodge star $*_4$ on Taub-NUT with any number of monopoles 
by specifying the orientation by the volume form as $e^0\wedge e^1\wedge e^2\wedge e^3$.
Then it is straightforward to check for instance the (anti-)selfdualtiy of $\Omega_I$ respectively $\Omega_I^\infty$  
noting that
\beq \label{eq:etaOmegaVierbeins}
\Omega_I = (1\pm *_4)\Big(\frac{V_I}{V}dU-dU_I\Big)\,.
\eeq

%%%%%%%%%%%%%%%%%%%%%%%%%%%%%%%%%%%%%%%%%%%%%%%%%%%%%%%%%%%%%%%%%%%%%%%%%%%%%%%%%%%%%%%%%%%%%%%%%%%%%%%
{
\providecommand{\href}[2]{#2}\begingroup\raggedright\endgroup
}
%\printindex
%\addcontentsline{toc}{chapter}{\numberline{}{Index}}
%%%%%%%%%%%%%%%%%%%%%%%%%%%%%%%%%%%%%%%%%%%%%%%%%%%%%%%%%%%%%%%%%%%%%%%%%%%%%%%%%%%%%%%%%%%%%%%%%%%%%%%

\end{document}